\documentclass{article}

\usepackage{arxiv}

\usepackage[utf8]{inputenc} 
\usepackage[T1]{fontenc}    
\usepackage{hyperref}       
\usepackage{url}            
\usepackage{booktabs}       
\usepackage{amsfonts}       
\usepackage{nicefrac}       
\usepackage{microtype}      
\usepackage{natbib}
\usepackage{doi}

\usepackage{graphicx}
\usepackage{float}
\usepackage{subcaption}
\usepackage{listings}
\usepackage{xcolor}
\usepackage{multirow}
\usepackage{etaremune}
\title{ADx3: A Collaborative Workflow for High-Quality Accessible Audio Description}

\author{
Lana Do\\
Northeastern University\\
San Jose, CA 95113\\
\texttt{do.ng@northeastern.edu}
\And
Shasta Ihorn\\
San Francisco, CA 94132\\
\texttt{sihorn@sfsu.edu}
\And
Charity Pitcher-Cooper\\
Smith-Kettlewell Eye Research Institute\\
San Francisco, CA 94115\\
\texttt{cpc@ski.org}
\And
Juvenal Francisco Barajas\\
San Francisco State University\\
San Francisco, CA 94132\\
\texttt{jbarajas8@mail.sfsu.edu}
\And
Gio Jung\\
San Francisco State University\\
San Francisco, CA 94132\\
\texttt{gjung1@sfsu.edu}
\And
Xuan Duy Anh Nguyen\\
San Francisco State University\\
San Francisco, CA 94132\\
\texttt{anguyen96@sfsu.edu}
\And
Sanjay Mirani\\
San Francisco State University\\
San Francisco, CA 94132\\
\texttt{smirani1@mail.sfsu.edu}
\And
Ilmi Yoon\\
Northeastern University\\
San Jose, CA 95113\\
\texttt{i.yoon@northeastern.edu}
}

\date{}

\renewcommand{\headeright}{}
\renewcommand{\undertitle}{}

\begin{document}
\maketitle

\begin{abstract}
	Audio description (AD) makes video content accessible to blind and low-vision (BLV) audiences, but producing high-quality descriptions is resource-intensive. Automated AD offers scalability, and prior studies show human-in-the-loop editing and user queries effectively improve narration. We introduce ADx3, a novel framework integrating these three modules: GenAD, upgrading baseline description generation with modern vision-language models (VLMs) guided by accessibility-informed prompting; RefineAD, supporting BLV and sighted users to view and edit drafts through an inclusive interface; and AdaptAD, enabling on-demand user queries. We evaluated GenAD in a study where seven accessibility specialists reviewed VLM-generated descriptions using professional guidelines. Findings show that with tailored prompting, VLMs produce “good” descriptions meeting basic standards, but “excellent” descriptions require human edits (RefineAD) and interaction (AdaptAD). ADx3 demonstrates collaborative workflows for accessible content creation, where components reinforce one another and enable continuous improvement: edits guide future baselines and user queries reveal gaps in AI-generated and human-authored descriptions.
\end{abstract}


\section{Introduction}

Digital video dominates education, entertainment, and communication, yet remains largely inaccessible for the estimated 285 million blind and low-vision (BLV) individuals worldwide. Audio description (AD)—spoken narration of visual elements—has been shown to substantially improve comprehension and engagement for BLV users, but production remains labor-intensive and rare outside professional contexts \cite{Schmeidler2000AddingDifference, Naraine2018ImpactsStudy}. Currently, AD remains largely unavailable on short-form, user-generated platforms such as YouTube, TikTok, and Instagram, which now represent the majority of digital video consumption, creating significant accessibility gaps that disproportionately affect BLV users’ access to contemporary digital culture.

Volunteer-driven systems such as YouDescribe \cite{YouDescribeYouDescribe.Https://www.youdescribe.org/}, Rescribe \cite{Pavel2020Rescribe:Descriptions} and LiveDescribe \cite{BranjeLiveDescribe:Description} provide streamlined authoring interfaces and collaborative editing support to lower barriers to AD creation and democratize description creation. These manual approaches, while valuable, cannot scale to match the volume of content produced daily across digital platforms. More recently, automated approaches using vision–language models (VLMs) demonstrate scalability but still fail to capture the contextual awareness, cohesive narration and timing precision that make AD effective.

Yuksel et al. developed a human-in-the-loop (HITL) approach that combined automatic generation with human editing. Their study with BLV users and novice describers showed that this method improved description quality while also lowering the barrier for new volunteers \cite{Yuksel2020Human-in-the-LoopUsers}. These results were notable even when relying on earlier captioning models. Prior research in machine translation (MT) also suggests that higher-quality outputs lead to a reduced editing workload and increased efficiency \cite{Green2013TheTranslation, LaubliAssessingEnvironment}, suggesting that stronger AI baselines in AD could likewise expand capacity. Building on this foundation, we advance baseline generation by moving beyond captioning pipelines such as Pythia combined with external OCR \cite{Stangl2023TheVideos}. We experiment with three modern vision–language models—Qwen2.5-VL \cite{Bai2025Qwen2.5-VLReport}, Gemini 1.5 Pro \cite{GoogleDeepMind2024GeminiReport}, and GPT-4o \cite{OpenAI2024GPT-4oCard}—that can generate visual descriptions and recognize on-screen text within a single model.

While HITL editing lowers the barrier for describers, other work has focused on giving BLV users more agency in the process, with interactive systems that let them request the details most relevant to their viewing goals. Stangl et al. demonstrated the value of interactive query tools alongside automated narration in studies with BLV participants \cite{Stangl2023TheVideos}, underscoring the importance of tailoring AD to user needs. Taken together, these studies suggest promise in automation, editing, and interactivity, yet no system has fully integrated them into a cohesive workflow.

We present ADx3, an end-to-end framework that brings together AI, human editors, and BLV audiences in a collaborative workflow. The system is structured around three modules: (1) GenAD leverages modern VLMs and prompting strategies to generate baseline descriptions; (2) RefineAD supports accessible, collaborative editing by describers; and (3) AdaptAD empowers BLV audiences to ask questions and request additional details, personalizing AD to their goals. This process is iterative, as edits and user interactions feed back into future prompting and fine-tuning. 

To evaluate ADx3, we asked two research questions (RQs):
\begin{enumerate}
\item \textbf{RQ1.} How well do modern VLMs generate baseline descriptions? More specifically, what are the strengths and weaknesses of AI-generated descriptions (GenAD) within our system?
\item \textbf{RQ2.} How does the integration of GenAD, RefineAD, and AdaptAD improve narration quality and user experience?
\end{enumerate}

We address these questions through an evaluation workshop in which seven accessibility experts assessed anonymized descriptions generated by Qwen2.5-VL, Gemini 1.5 Pro, and GPT-4o using established AD guidelines. Prior work established the value of HITL edits and user on-demand queries through extensive studies with BLV participants \cite{Yuksel2020Human-in-the-LoopUsers, Stangl2023TheVideos}. To complement this, our study engages accessibility experts to evaluate the upgraded baseline generation (GenAD), an assessment requiring both accessibility knowledge and the ability to verify visual detail. We further provide a formative demonstration of how RefineAD and AdaptAD extend this baseline into a cohesive system. Findings show that while VLMs can produce fluent and detailed drafts, targeted refinement remains essential to elevate an adequate draft into a high quality one, and user-driven interaction is critical for adapting descriptions to diverse contexts. Our contributions are: (1) an evaluation of multiple VLMs for generating baseline descriptions, identifying their strengths and limitations and (2) a unified system design that embeds both HITL editing and on-demand queries into a scalable pipeline for high-quality accessible AD.

\section{Related Work}

\subsection{Accessibility Guidelines and Volunteer-driven Platforms for Audio Description }

Audio description has long been required in film and broadcast, supported by the emergence of professional frameworks and industry guidelines. Web Content Accessibility Guidelines 2.0 Level AA mandates AD for all prerecorded synchronized media \cite{2008WorldPrerecorded.}. The Described and Captioned Media Program (DCMP) provides widely cited guidance for educational and instructional contexts, emphasizing precision and consistency in learning environments \cite{2024DescribedDCMP}. The National Center for Accessible Media (NCAM) at WGBH Educational Foundation, a Boston-based public media organization, pioneered AD research and developed standards for broadcast and film, establishing conventions of timing, style, and integration that remain influential today \cite{NationalCenterforAccessibleMedia2017AccessibleGuidelines}. In the commercial sector, companies such as 3Play Media have codified AD workflows as part of broader accessibility services, embedding description, captioning, and subtitling into professional production pipelines. Collectively, these sources have established benchmarks that shape practice for both professional describers and volunteer contributors, influencing expectations of quality, style, and timing across domains.

With the rapid growth of digital and user-generated video, the demand for AD has expanded well beyond cinema and television into platforms such as YouTube, TikTok, and Instagram, where content is produced at massive scale and on short timelines \cite{Schmeidler2000AddingDifference, Naraine2018ImpactsStudy}. In response, early authoring tools and crowdsourcing platforms sought to broaden access by lowering barriers for non-professionals, adapting professional accessibility guidelines to enable community-based AD creation. LiveDescribe provided non-experts with a simple interface to record and align AD with video timelines \cite{BranjeLiveDescribe:Description}, while Rescribe streamlined authoring by integrating scripting, timing alignment, and audio editing into one collaborative tool \cite{Pavel2020Rescribe:Descriptions}. YouDescribe demonstrated the feasibility of crowdsourced AD, allowing volunteers to add descriptions to YouTube videos on request by BLV users \cite{YouDescribeYouDescribe.Https://www.youdescribe.org/}. These approaches expanded AD availability, but their reliance on volunteer labor made them difficult to sustain at scale.

While volunteer-driven platforms broadened availability, their reliance on manual labor made it difficult to keep pace with the volume of online content and sustain consistent quality. Meeting this demand requires more advanced automation that can extend the reach of existing guidelines and community contributions while ensuring alignment with accessibility standards.

\subsection{Automating AD and Human-in-the-Loop Refinement}

Advances in multimodal learning have expanded the potential for automated AD. Systems like Tiresias aligned event-based AD with salience cues and audio using hierarchical attention \cite{Wang2021TowardVideos}. Later models such as BLIP-2 \cite{Li2023BLIP-2:Models} combined frozen vision and language models through an adapter, enabling generative tasks like captioning and VQA but limiting deeper temporal reasoning. Recent pipelines leverage these models for AD—for example, LLM-AD used GPT-4V with scene segmentation and character tracking \cite{Chu2024LLM-AD:System}, while MMAD combined CLIP with LLaMA2 to integrate subtitles and sound cues for narrative consistency \cite{Ye2024MMAD:Description}. However, automated AD alone often falls short: systems misidentify characters or insert hallucinated details \cite{Bergin2025AutomatingAlgorithms}, produce verbose or redundant captions that confuse rather than aid \cite{Wang2021TowardVideos}, and score well on captioning metrics yet misalign with human judgments of AD quality \cite{Rohrbach2017MovieDescription}.

To improve clarity and usability, prior work explored human-in-the-loop (HITL) systems where humans edit AI-generated drafts. Yuksel et al. showed this approach produced descriptions BLV users rated higher in quality and comprehension, while also reducing describer workload \cite{Yuksel2020Human-in-the-LoopUsers}. However, that study involved only novice describers, focused on dialogue-free videos, and relied on captioning models (e.g., Microsoft Video Insight, Pythia) whose outputs were rated “neutral” in accuracy. Even so, participants reported reduced effort compared to writing from scratch, showing that even imperfect baselines can meaningfully lower workload. 

Research in machine translation further supports this point: professional translators editing high-quality neural MT outputs made fewer changes and worked significantly faster than when revising weaker outputs \cite{Green2013TheTranslation, LaubliAssessingEnvironment}. These findings suggest that more reliable AI-generated baselines in AD could likewise reduce cognitive and editing effort, expanding capacity for accessible description creation. We upgrade baseline generation from image captioning models to three state-of-the-art VLMs, including Qwen2.5-VL \cite{Bai2025Qwen2.5-VLReport}, Gemini 1.5 Pro \cite{GoogleDeepMind2024GeminiReport}, and GPT-4o \cite{OpenAI2024GPT-4oCard}, using accessibility-guided prompting under identical pipeline conditions. The prompting strategies were iteratively refined with feedback from an experienced consultant, who has experience working with BLV users and training novice describers, allowing our outputs to steadily improve in clarity and alignment with professional practice. The resulting descriptions are then evaluated in a study with accessibility specialists across diverse video types, including dialogue-heavy and text-heavy content. This work extends prior BLV-centered studies with novice describers by analyzing which aspects of AI-generated descriptions specialists seek to improve, and how they use the system to make those improvements.

\subsection{Interactivity and Human-in-the-Loop Systems}

While HITL refinement improves clarity and correctness, BLV audiences also have diverse and context-dependent needs that cannot be met by fixed narration alone. Studies show that user preferences vary with viewing goals and genres of the content \cite{JiangCrescentiaJungMahikaPhutaneItsScenarios}. For instance, in social or family videos, some users prefer richer character descriptions and emotional context to better follow interpersonal dynamics. In contrast, during educational videos, users often favor more concise and focused narration that highlights key instructional content without unnecessary detail. A truly accessible system must allow BLV users to adjust the type and level of detail to match their individual goals and viewing contexts.

A growing body of work emphasizes interactive systems that give BLV users greater agency in how they access visual information. DescribeNow \cite{Cheema2024DescribeIndividuals} let viewers adjust narration style, for instance, switching between detailed descriptions and concise summaries, while ShortScribe \cite{VanDaele2024MakingSummaries} enabled toggling across levels of detail, from shot-by-shot narration to long-form summaries and high-level overviews. SPICA \cite{Ning2024SPICA:Viewers} extended interactivity by enabling temporal navigation through frame captions and spatial exploration of objects within key frames. Together, these systems illustrate how interactivity moves beyond static, one-size-fits-all narration, giving users meaningful control over the level, type, and timing of descriptive information. Complementary studies have also shown the benefits of pairing baseline narration with interactive agents. Bodi et al. \cite{Bodi2021AutomatedUsers} and Stangl et al. \cite{Stangl2023TheVideos} found that BLV users reported greater comprehension, enjoyment, and satisfaction when they can request supplemental detail or ask specific questions. 

Existing studies show that automation alone is insufficient, baseline drafts need refinement, and narration must adapt to individual goals. These components remain fragmented across separate systems. In this work, these elements are integrated in a collaborative workflow: GenAD, which generates baseline descriptions through targeted prompting strategies with modern VLMs; RefineAD, which supports clarity and consistency through human editing; and most importantly, AdaptAD, which gives BLV users direct agency over when and how descriptions are delivered. To clarify where AdaptAD can add the most value, our study complements prior BLV-centered evaluations \cite{Bodi2021AutomatedUsers, Stangl2023TheVideos} by engaging accessibility experts to assess generated AD baselines (GenAD), thereby surfacing patterns of omission and ambiguity. Beyond serving immediate needs, AdaptAD also creates a feedback channel: by analyzing patterns in what users ask, whether requests cluster around character identification, spatial relationships, or frequent text-on-screen events, the system can reveal systematic gaps in AI descriptions and human edits alike.

\section{System}

\begin{figure}[h]
  \centering
  \includegraphics[width=\linewidth]{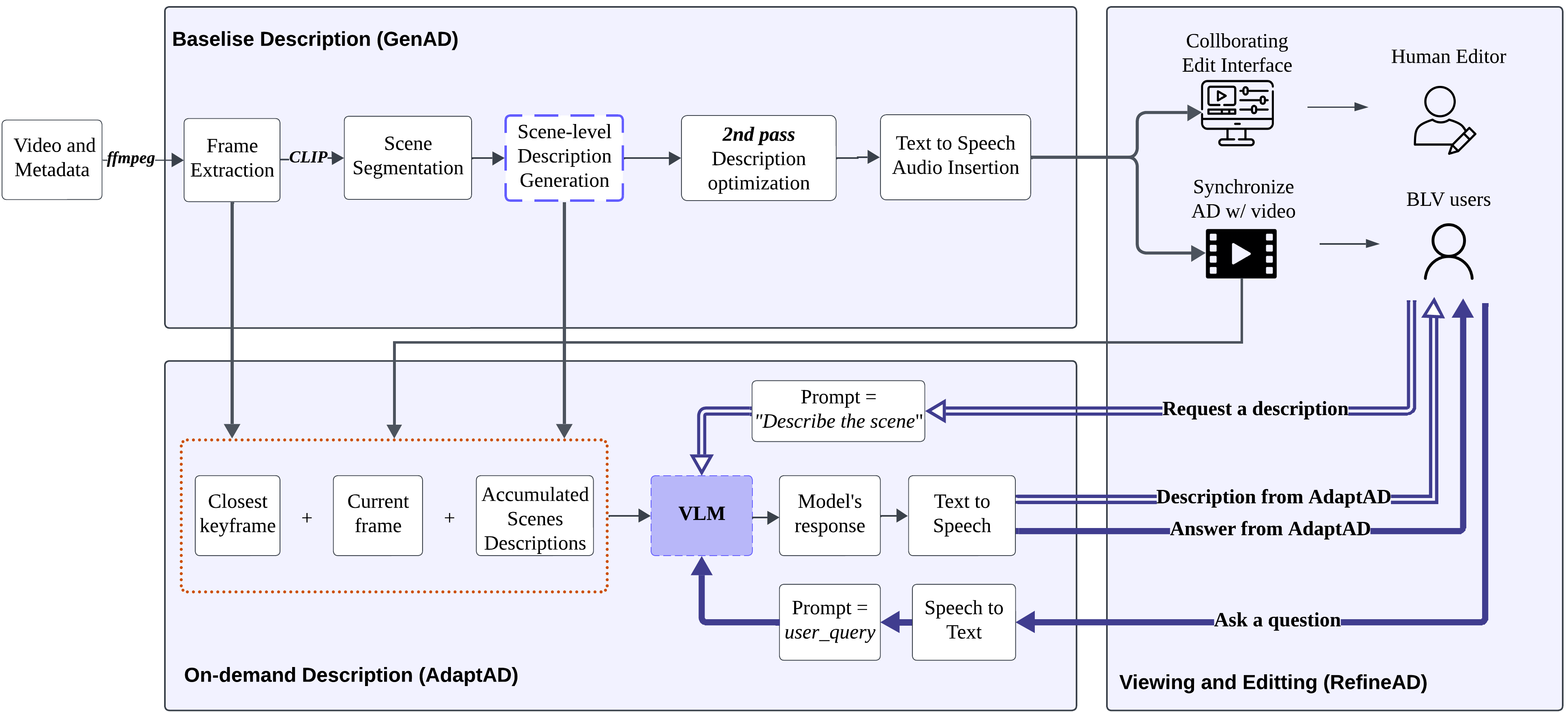}
  \caption{ADx3 system architecture showing three integrated modules: GenAD for baseline generation, RefineAD for human editing, and AdaptAD for user-driven interaction with synchronized audio description}
  \label{system-overview}
\end{figure}

Fig.~\ref{system-overview} illustrates the complete ADx3 system with three modules: GenAD for initial description generation and AdaptAD for user interaction, both powered by VLMs, and RefineAD providing the human editing layer. 

Since GenAD and AdaptAD depend on VLMs, selecting appropriate models is critical. Video-MME is a large-scale benchmark that evaluates 50 VLMs on multimodal reasoning, temporal grounding, and consistency across diverse video tasks \cite{Fu2024Video-MME:Analysis}. On its latest leaderboard, Gemini ranked first overall, Qwen-VL ranked fifth, and GPT ranked sixth. We selected these models not only because they performed among the top systems, but also because they are accessible for research: Qwen2.5-VL is open-source on Hugging Face, while Gemini 1.5 Pro and GPT-4o are available via commercial APIs. To enable a controlled comparison, our pipeline is run with each of the three models while keeping all other components identical.

\subsection{Baseline Description Generation: GenAD}
The baseline description is designed to convey critical visual details of a video to BLV audiences, including prominent objects, characters, their movements, and interactions, visible text, and environmental cues such as setting and attire. Our system supports two delivery modes: \textit{Inline} narration, which inserts descriptions into natural pauses in dialogue, and \textit{Extended} narration, which briefly pauses playback to deliver additional detail. Traditional AD practices have largely focused on inline delivery, reflecting both broadcast constraints and the preference to avoid interrupting content. However, inline descriptions alone are often insufficient in dialogue-dense videos or content that requires precision, such as educational materials \cite{FundamentalsDescription}. In these cases, extended narration can provide critical information that would otherwise be omitted. The system is designed to support both modes technologically, enabling flexible delivery strategies that balance fidelity, timing, and accessibility across diverse video types.

\subsubsection{Video Input and Scene Segmentation}
The workflow begins by retrieving the target video using \texttt{yt-dlp}, a command-line utility for downloading videos and associated metadata from YouTube. Metadata includes the video title, description, category, and subtitles when available. Video frames are first extracted using \texttt{ffmpeg}, a command-line multimedia framework \cite{TomarSuramya2006ConvertingFFmpeg}. To segment the video into coherent scenes, we use OpenCLIP’s visual encoder to extract frame-level embeddings \cite{Schuhmann2022LAION-5B:Models}. Cosine similarity is computed between consecutive frames, and a scene boundary is marked when the similarity drops below a predefined threshold. This method enables the system to group visually consistent segments and helps reduce redundancy in the generated descriptions.

\subsubsection{Scene-level Description Generation}

\begin{figure}[H]
  \centering
  \includegraphics[width=0.7\linewidth]{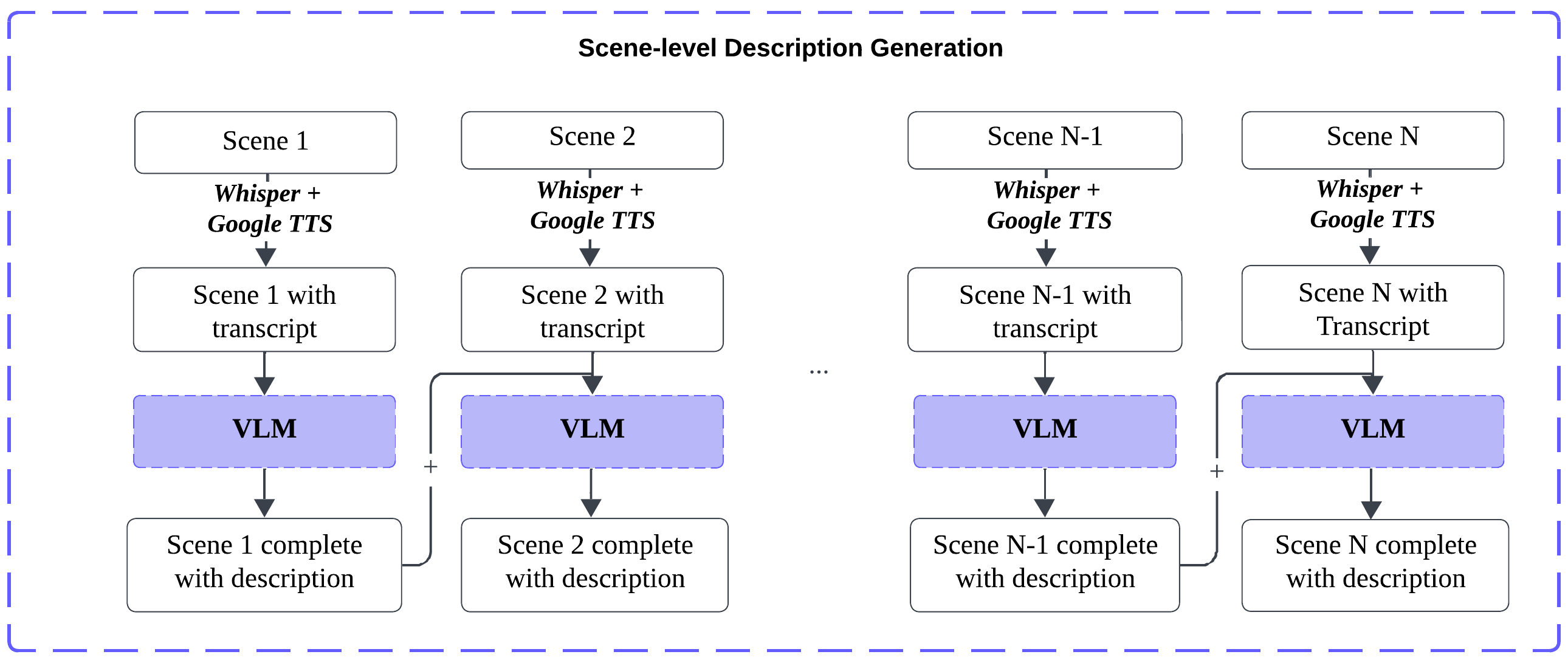}
  \caption{Scene-level description generation using contextual prompting from audio transcripts and accumulated prior scene descriptions.}
  \label{scene-generation}
\end{figure}

For each segmented scene, early experiments prompting VLMs with only the scene's frames produced outputs that looked more like captions than coherent audio description. They were verbose, repeated dialogue, and often vague about characters or context, issues also documented in prior research and guidelines \cite{Sarhan2023UnderstandingImages, Stangl2023TheVideos, 2024DescribedDCMP}. When we shared these drafts with an experienced audio describer who also trains volunteers, they observed that the VLM lacked contextual references and cohesive flow, and that its style resembled trainee describers who \textit{“try to do too much”} by narrating every visible detail, including information that can be inferred from the soundtrack.

To address these issues, we introduced two key strategies. First, persona prompting was added to cast the model as a \textit{“professional describer.”} This was reinforced with a guided self-check step, where the model was presented with a short set of audio description guidelines commonly recommended to human describers and asked to confirm understanding before generating. The guidelines emphasized conciseness, factual accuracy, and tone matching, while discouraging over-description, interpretation, and premature spoilers, among other practices (the full prompt is provided in the Appendix) \cite{YouDescribeYouDescribe.Https://www.youdescribe.org/}. Second, we improved contextual grounding by supplementing the prompt with the video’s metadata (titles and summaries), audio transcripts of the scene, and accumulated descriptions from prior scenes. This provided continuity across segments and avoided vague or generic phrasing. 

Accurate audio transcription was critical for both contextual generation and delivery optimization: without reliable transcripts, the model could not generate precise descriptions, and without accurate timestamps, the system could not optimize inline delivery. To ensure both, we used an ensemble approach where Whisper produced word-level timestamps via Dynamic Time Warping, and Google Speech-to-Text served as a cross-check to mitigate hallucinations, with only reliable material retained. An overview of this scene-level generation pipeline is illustrated in Figure \ref{scene-generation}.

Finally, the prompt specified the output format, requiring two labeled event types for each scene—\textit{Text-on-Screen} and \textit{Visual}, and to include the explicit start time of each event, providing alignment with the video timeline and greater usability for subsequent optimization.

\subsubsection{Description Optimization}

Even with improved generation, the outputs were often too verbose for seamless delivery. Guided by continued feedback from our consultant, we refined the prompts and narration strategies to better align with BLV user preferences for concise narration that minimizes interruptions \cite{2024DescribedDCMP, NationalCenterforAccessibleMedia2017AccessibleGuidelines, 3PlayMedia2020AudioGuidelines}. To address this, we applied a second pass of prompt-based optimization. Inline optimization condenses narration to fit available dialogue gaps, with retry prompts used to further shorten descriptions while prioritizing critical details. Extended narration is reserved for cases where inline delivery is impossible but essential information would otherwise be omitted. Here, a filtering prompt casts the model as an \textit{“accessibility expert,”} instructing it to mark descriptions as \textit{“necessary”} only if they convey silent actions, novel visual elements, or character clarification. For How-to videos, where fragmented narration of \textit{Text-on-Screen} and \textit{Visual} actions can be distracting, a merging prompt combines both into a single coherent sentence while preserving precise measurements. This strategy reflects best practices in AD for procedural content, reinforced through iterative consultant feedback. Through accurate transcription, guideline-informed prompting, contextual grounding, and iterative refinements, GenAD transforms initial VLM output into concise, coherent narration that reflects professional AD practice and meets BLV audiences’ expectations for clarity and quality.

\subsubsection {Illustrative Improvements from Prompting Design}

\begin{figure}[H]
    \centering

    \begin{subfigure}[t]{0.6\linewidth}
        \centering
        \includegraphics[width=\linewidth]{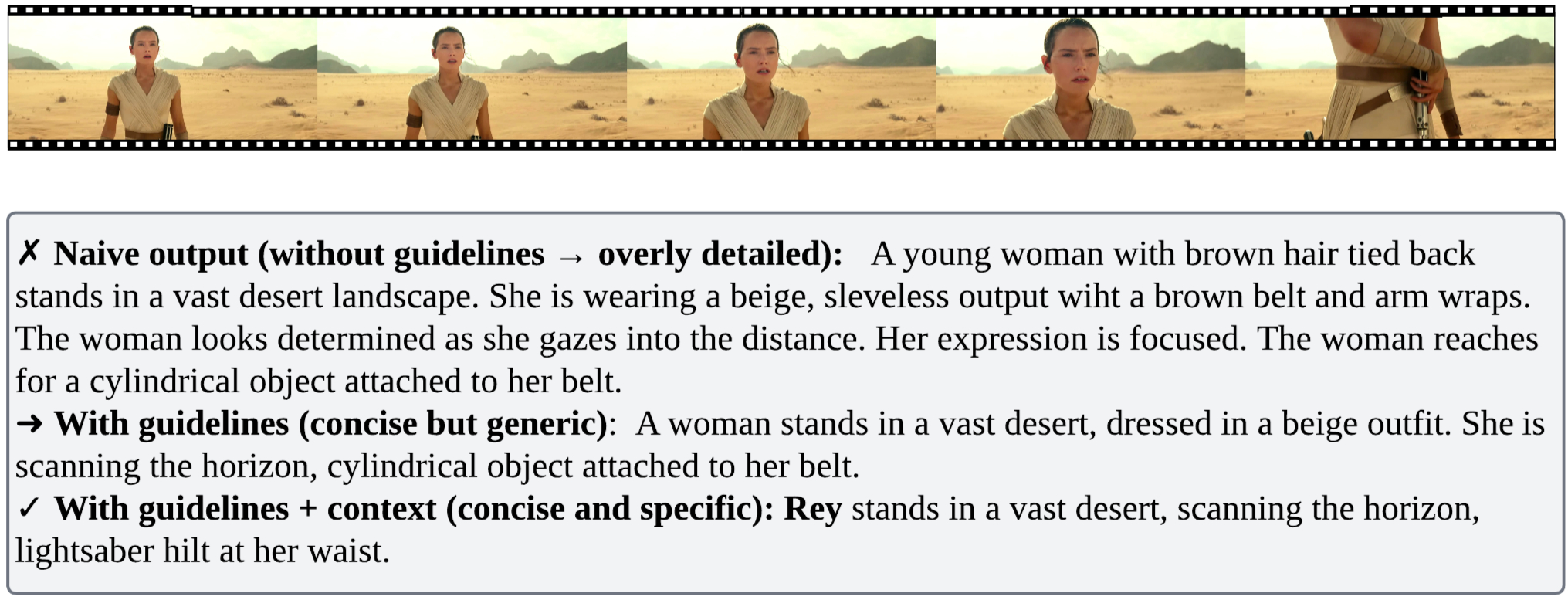}
        \caption{Star Wars Trailer – progression from simple VLM prompting to guideline- and context-informed prompting.}
        \label{starwars-output}
    \end{subfigure}

    \vspace{1em}

    \begin{subfigure}[t]{0.6\linewidth}
        \centering
        \includegraphics[width=\linewidth]{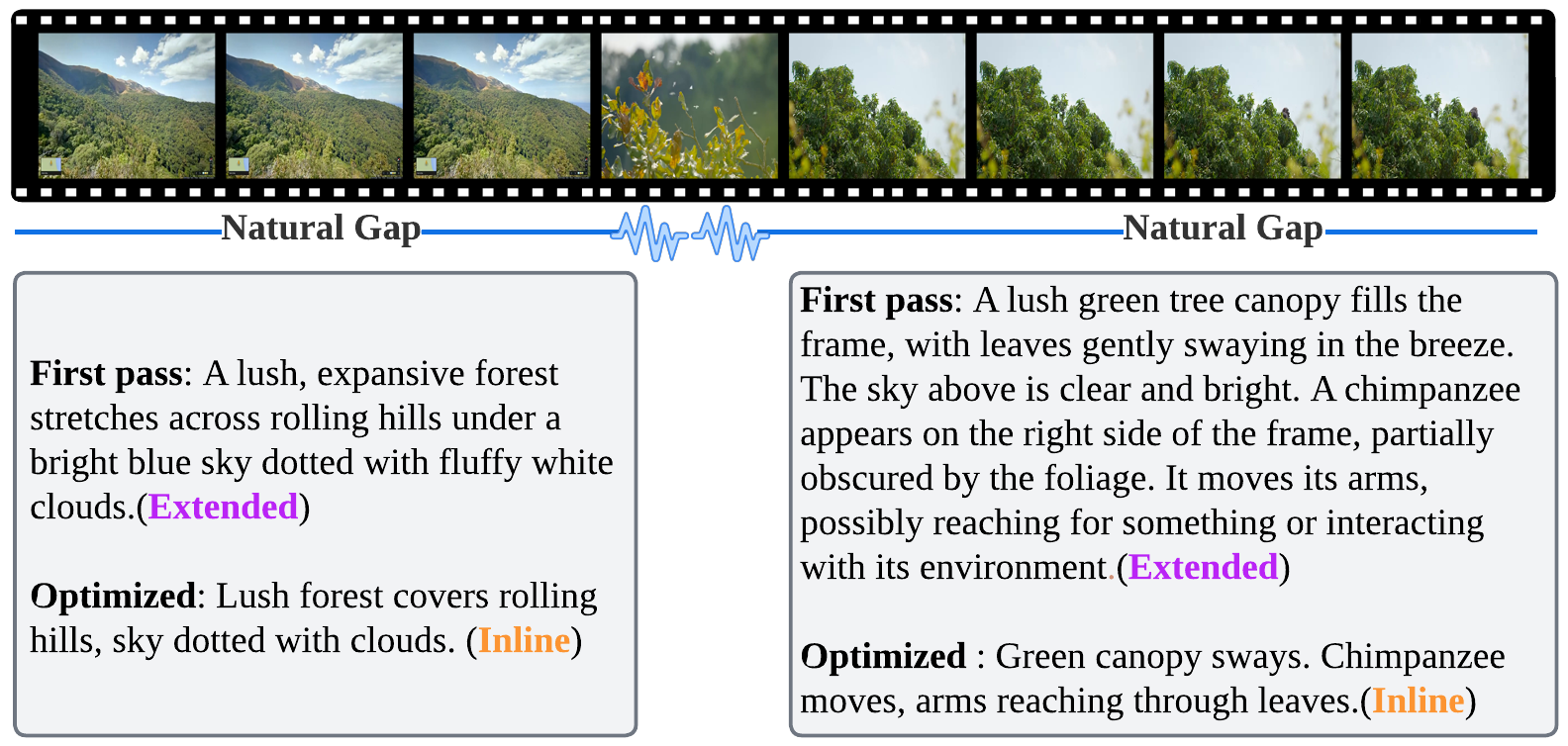}
        \caption{Jane Goodall clip – optimization condenses long drafts into descriptions that fit natural pauses.}
        \label{jane-output}
    \end{subfigure}

    \vspace{1em}

    \begin{subfigure}[t]{0.6\linewidth}
        \centering
        \includegraphics[width=\linewidth]{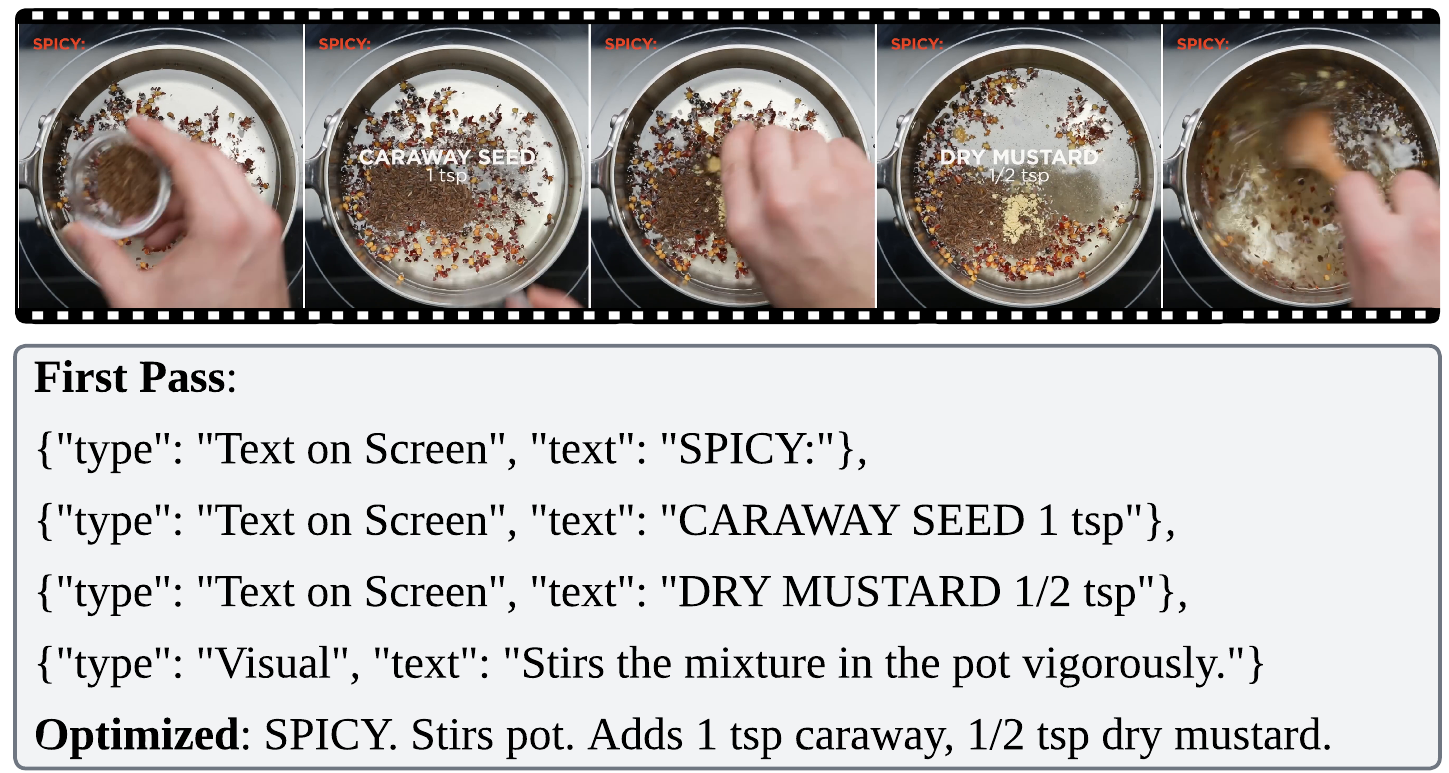}
        \caption{3 Ways to Make Pickles – optimization merges text and visual cues into one concise description.}
        \label{pickles-output}
    \end{subfigure}

    \caption{Examples of prompting and optimization improvements across different videos: (a) refined prompting with guidelines and context, (b) condensing verbose drafts into natural pauses, and (c) merging text and visual cues into fluent descriptions.}
    \label{prompting-optimization}
\end{figure}

We highlight several qualitative improvements observed during iterative design in Fig. \ref{prompting-optimization}. With scene-level generation, guideline-based and contextual prompting produced more specific narration: raw VLM output described \textit{“a woman stands in the desert”}, while the refined version identified her as \textit{“Rey”} (Fig.\ref{starwars-output}). With description optimization, prompts reduced verbosity and improved coherence. Fig.\ref{jane-output} shows how an overlong draft that would have required an extended track was condensed into an inline description that fit natural pauses, maintaining flow without losing detail. Fig.~\ref{pickles-output} demonstrates how merging prompts combined on-screen text with simultaneous actions into a single fluent line, mirroring how a human describer might naturally convey the scene. These formative observations demonstrated that our strategies mitigated common shortcomings of using VLMs without accessibility-guided prompting, providing the basis for a more comprehensive and nuanced expert evaluation in the next section.

\subsubsection{Text to Speech Audio Insertion} 

Optimized descriptions are then converted to speech and aligned with the video timeline. Inline narration is inserted during natural pauses, while extended narration triggers a brief pause in playback to deliver essential details without overlap. This insertion strategy ensures that descriptions enhance rather than interfere with the existing audio. The finalized text is synthesized into speech using Google Text to Speech (TTS). To help distinguish content types, \textit{Visual} descriptions are voiced with a female voice and \textit{Text-on-Screen} entries with a male voice—an approach informed by our consultant, who noted that repeatedly hearing the phrase \textit{“text on screen”} can become bothersome for BLV users. The audio clips are then synchronized with the original video and surfaced in the interactive interface for both playback and refinement.

\subsection{Human in the Loop (HITL) Workflow}
\begin{figure}[H]
    \centering
    \includegraphics[width=0.65\linewidth]{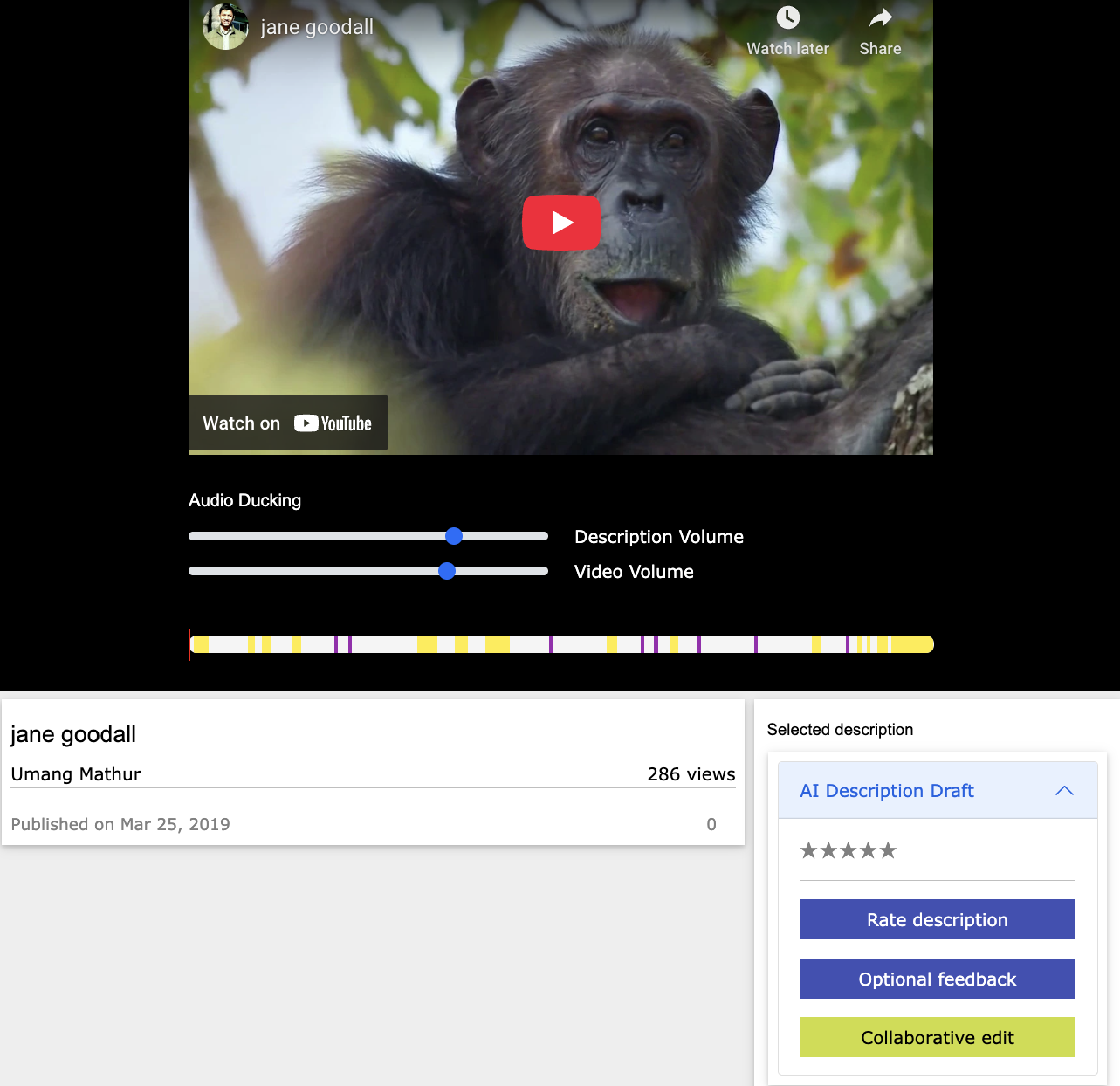}
    \caption{Users preview the AI-generated draft, with inline and extended AD tracks in timeline bar. Buttons allow rating, optional feedback, and (by default) enabling collaborative editing to open the editor.}
    \label{view}
\end{figure}

The interface to view the generated descriptions is presented in Fig.~\ref{view}. To ensure accessibility, the platform implements Web Accessibility Initiative – Accessible Rich Internet Applications (WAI-ARIA), a technical specification that provides semantic roles, states, and properties to make interactive web content perceivable to assistive technologies such as screen readers (e.g., JAWS, NVDA, VoiceOver) \cite{WAI-ARIAW3C}. Combined with high-contrast mode and full keyboard navigation, these foundations allow the platform to remain inclusive across devices and user needs.

Within the viewing interface, inline narration appears as yellow segments embedded in the timeline, while extended narration is indicated by purple markers that briefly pause playback for longer details. This color coding allows sighted users to distinguish narration types at a glance. Playback is further supported by audio ducking controls to balance narration with original audio, full keyboard navigation. The AI-generated description can be rated and commented on, establishing a feedback loop that identifies useful drafts and highlights those requiring refinement. 

\begin{figure}[H]
    \centering
    \includegraphics[width=0.85\linewidth]{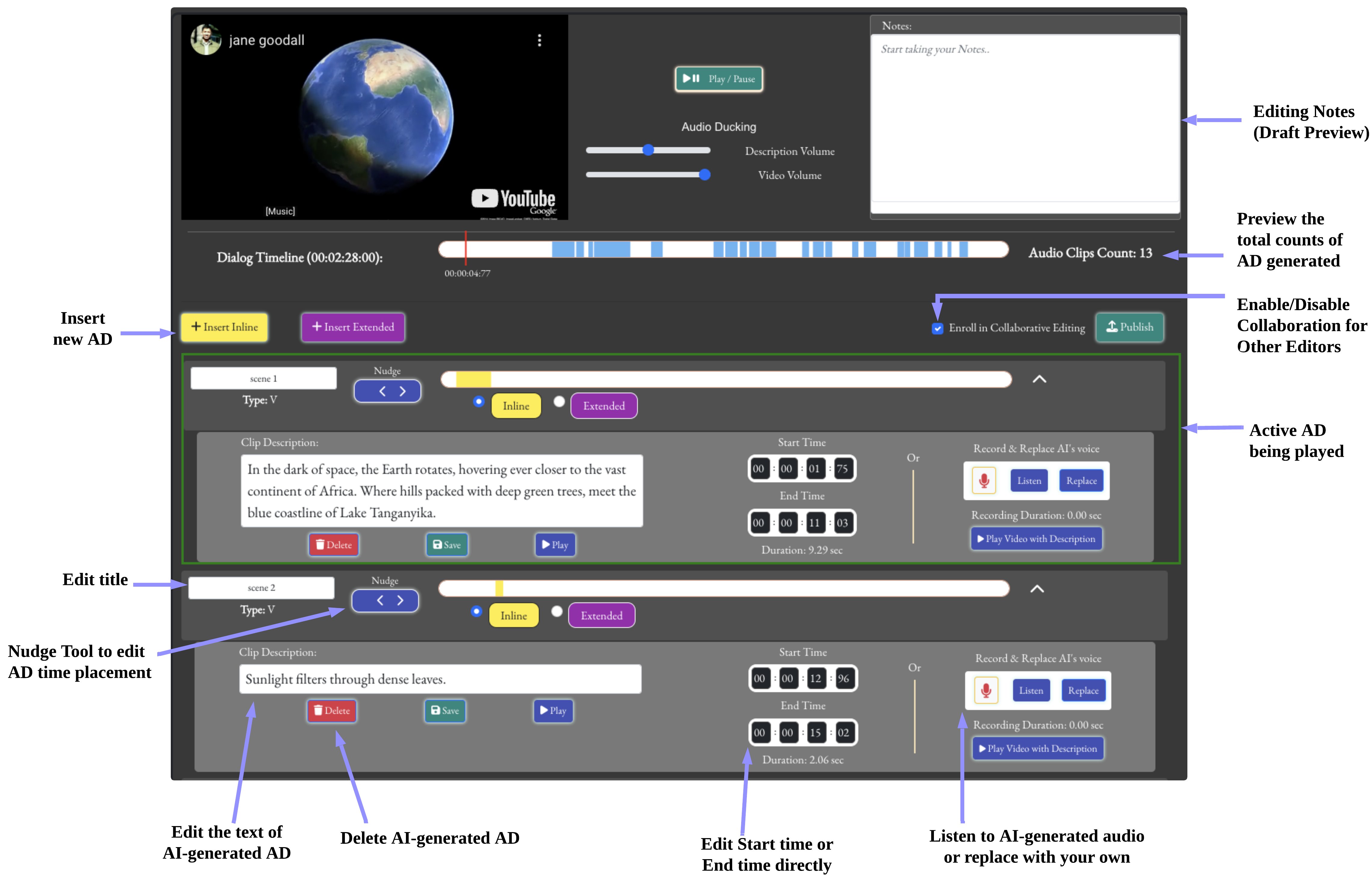}
    \caption{Editors refine AD drafts with features such as text revision, delivery style switching, track alignment, timestamp editing with the nudging tool, and adding/removing tracks. Accessibility support (screen reader, high contrast, keyboard navigation) enables BLV users to participate directly.}
    \label{colab}
\end{figure}

For more substantive and detailed edits, users click the yellow “Collaborative Edit” button to open the editing interface (Fig.~\ref{colab}). Draft descriptions are pre-labeled by type (Visual or Text-on-Screen) and delivery style (inline or extended). As the video plays with the AI draft synchronized, the active segment is outlined with a green bounding box. Editors can also use the notes section to plan revisions. Within the interface, users can revise text, switch delivery style, align segments, add new tracks, and remove unwanted AI-generated ones. Timing can be adjusted directly in the timestamp box or fine-tuned with the \textbf{nudging tool}, which shifts cues left or right with frame-level precision. Designed with the same accessibility features as the viewing interface, the editing interface empowers both BLV users and sighted volunteers to correct inaccuracies and improve AD directly.

Synthetic voice tracks may also be replaced with editor-recorded narration for a more engaging and relatable delivery. Research shows that blind and visually impaired audiences consistently prefer human narration for its clarity and emotional depth, particularly in dramatic or fast-paced content where synthetic voices can feel flat or detached \cite{Fernandez-Torne2015TheCatalan, Walczak2018VocalPresence, RNIBRNIB}. This feature reflects a human-centered design, balancing the scalability of synthetic speech with the authenticity and expressiveness of recorded narration.

Before publication, authors can choose to opt in to \textit{Enable Collaborative Editing}. This setting is enabled by default only for the initial AI-generated draft to encourage human review. Once any edits have been made, collaborative editing becomes optional: describers may keep it on to allow additional editors, or turn it off so that only the original author can edit the draft. This model allows the editor to invite collaboration, aligning with prior work such as Viscene showing that collaboration between novice describers and experts—whether sighted professionals or BLV users with lived expertise—can improve narration quality while reducing workload for all contributors \cite{Natalie2020ViScene:Videos}. All progress is automatically saved, and describers can return at any time to make further edits or unpublish if needed.

\begin{figure}[H]
    \centering
    \begin{subfigure}[t]{0.325\linewidth}
        \centering
        \includegraphics[width=\linewidth]{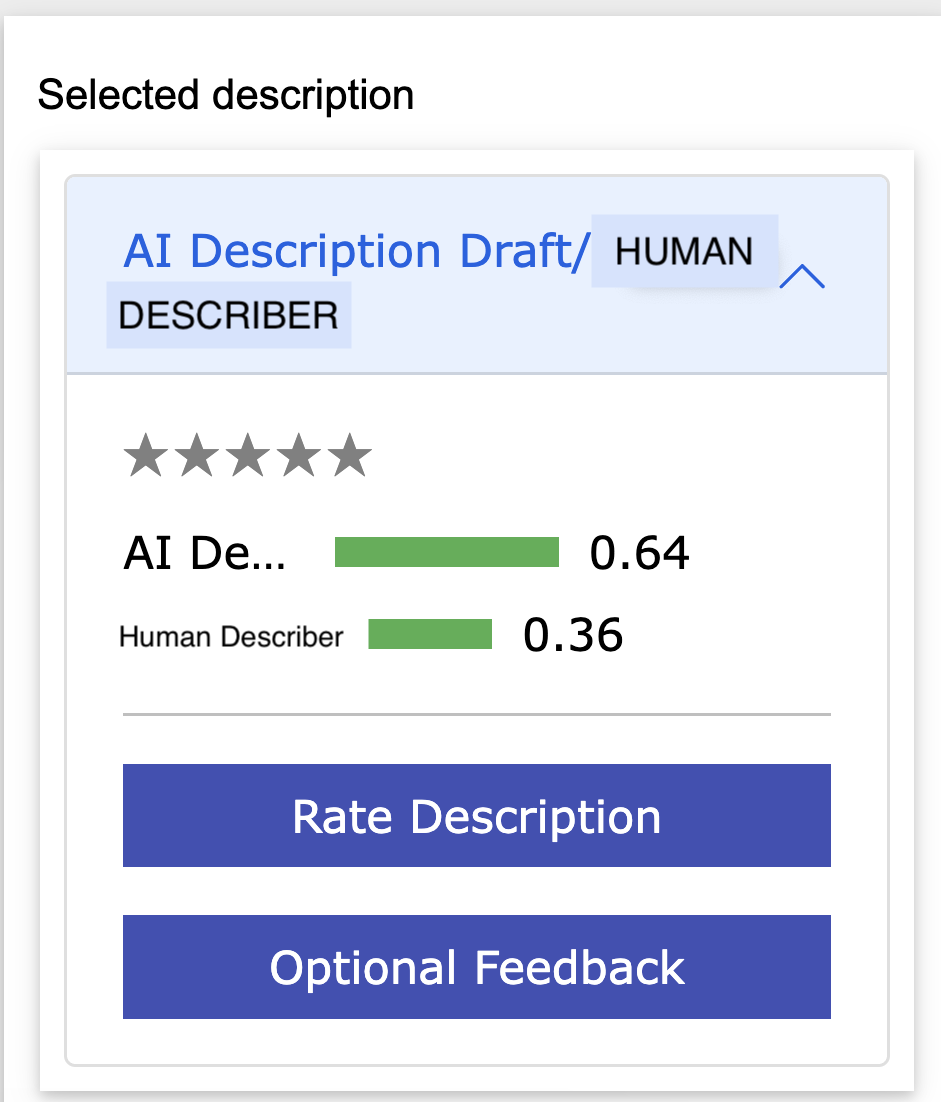}
        \caption{AI–human contribution breakdown with collaborative editing disabled. The option to invite others is not shown.}
        \label{fig:ai-human-contribution}
    \end{subfigure}
    \hspace{0.04\linewidth}
    \begin{subfigure}[t]{0.28\linewidth}
        \centering
        \includegraphics[width=\linewidth]{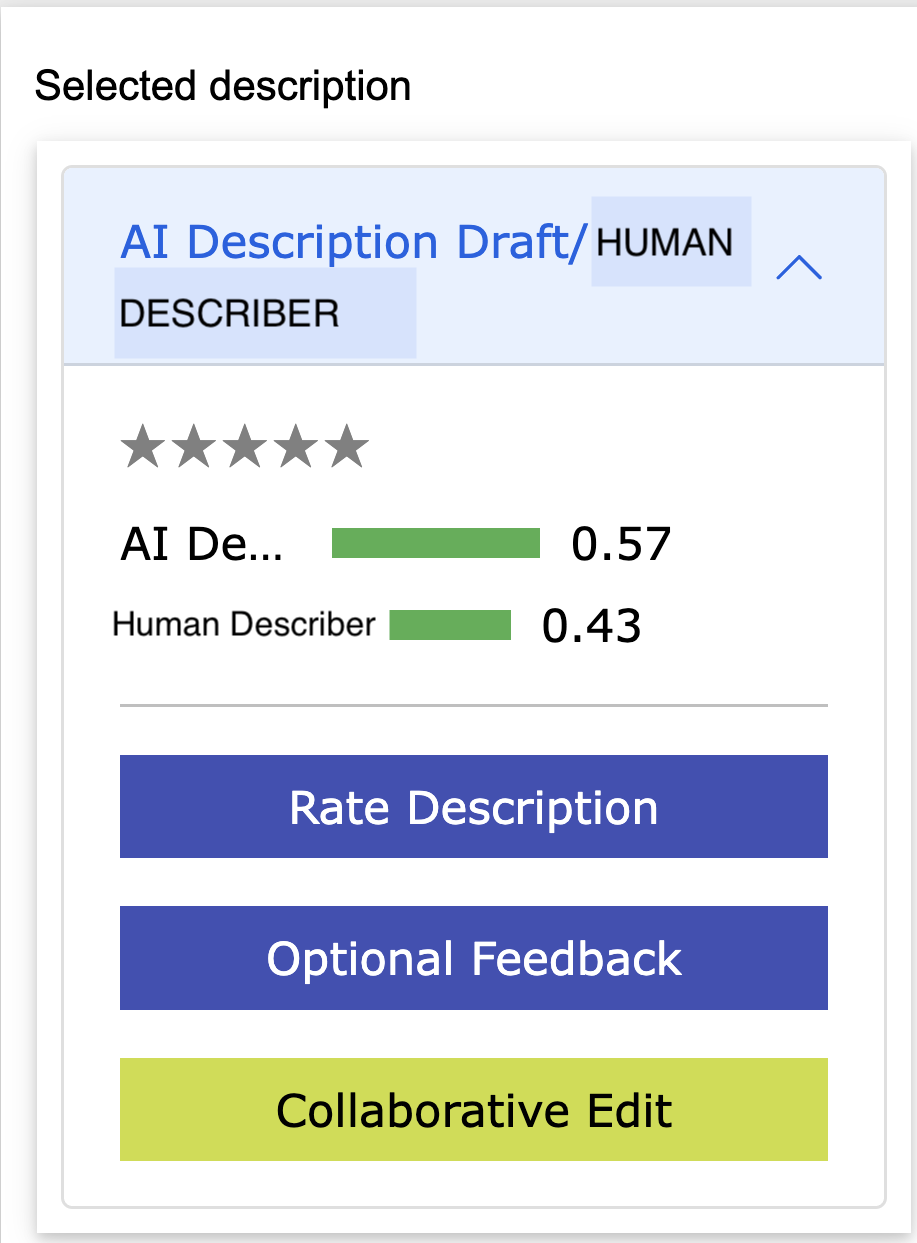}
        \caption{AI–human contribution breakdown with collaborative editing enabled, allowing further contributions.}
        \label{fig:colab-enabled}
    \end{subfigure}
    \hspace{0.04\linewidth}
    \begin{subfigure}[t]{0.29\linewidth}
        \centering
        \includegraphics[width=\linewidth]{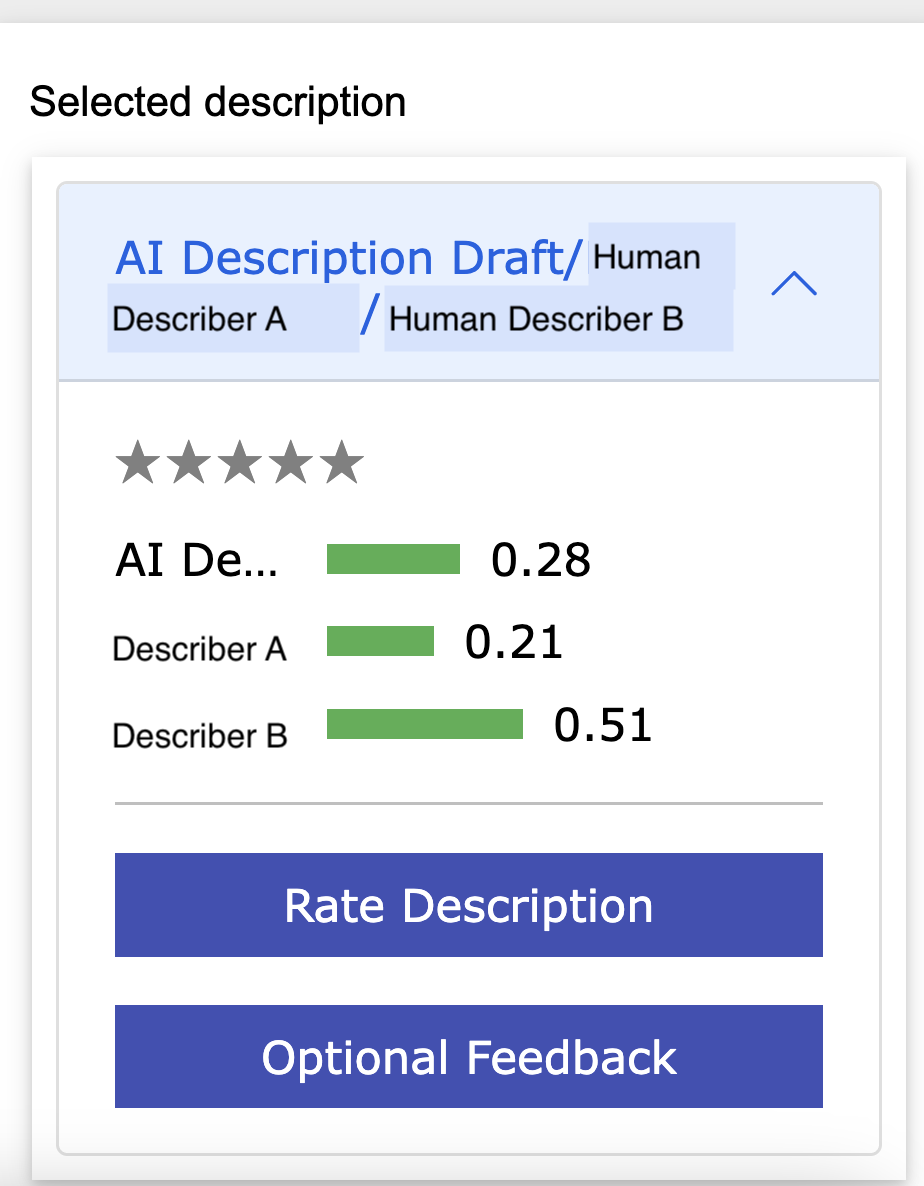}
        \caption{Collaborative editing view with contributions from multiple human editors on one draft.}
        \label{fig:multi-describer-ui}
    \end{subfigure}
    \caption{Authoring interface features: (a) shows AI–human contribution breakdown when collaborative editing is disabled; (b) shows the breakdown when collaborative editing is enabled, allowing additional contributors; and (c) shows the collaborative editing view with multiple human editors working on a single draft.}
    \label{contribution}
\end{figure}

Finalized description tracks are published showing the proportion of contributions from AI-human and also between different authors. All revisions are logged automatically (Fig.~\ref{contribution}), ensuring accountability and supporting iterative improvement. Contribution is quantified using an edit-distance method: each revision’s Levenshtein distance from the prior version is normalized by content length to assign proportional credit. This approach enables multi-author contribution and captures both the extent and sequence of human involvement. 

By providing editable drafts rather than requiring descriptions to be created from scratch, the system improves the quality of automated AD through structured human oversight, lowers barriers to participation, and broadens who can contribute. AI drafts serve as a starting point that human editors incrementally refine, supporting the scalability of AD while maintaining both quality and accountability. As draft quality improves, fewer edits are needed, reducing effort for editors. The interface is designed for ease of use, with features such as quick timing adjustments that simplify tasks that are traditionally complex. Through the integration of VLMs with accessibility-guided prompting and a feature-rich interface, the system advances accessibility efforts by delivering not only a greater quantity of audio descriptions for BLV users but also higher-quality narration shaped through editorial refinement.

\subsection{On-Demand Description: AdaptAD}
AdaptAD enables users to request additional information about a video at any point during playback. When activated, the video pauses automatically, allowing users to explore the visual content without missing subsequent action. To generate accurate answers, AdaptAD analyzes the current paused frame in combination with the nearest keyframe, especially when the paused frame may be blurred or visually incomplete. These frames, together with the scene transcript and previously generated descriptions outputs from GenAD, are passed to the VLM to produce responses that reflect the current visual state while maintaining narrative context. 

From consultation with our accessibility expert and prior studies with BLV users \cite{Stangl2023TheVideos}, we incorporated BLV users’ preference for concise, one-sentence responses rather than verbose outputs often produced by VLMs. Prompting in AdaptAD is designed to reflect this preference, guiding the model to generate contextually rich but concise answers that remain focused on the user’s query and highlight the most relevant details.

\subsubsection{Instant Description Request}
By pressing \textit{Option + D}, users can request an instant description of what is currently visible in the video. The resulting description is converted to speech using Google TTS and read aloud. This feature gives BLV users immediate access to visual details that may not be included in the original audio description.

\subsubsection{Question Answering}
By pressing \textit{Option + Q}, users can ask a spoken question about the paused video frame. The query is transcribed using Google STT (Speech-to-Text) and then passed to the VLM along with the relevant visual and contextual inputs described earlier. The model's response is then converted to speech via TTS and played back to the user. This feature enables users to actively explore and clarify specific aspects of the scene, supporting greater engagement and autonomy.

\subsubsection{Preliminary Observations}

In combination with GenAD, AdaptAD applies similar prompting strategies—encouraging VLMs to generate concise responses, as recommended by our consultant and prior work—and incorporates additional context such as scene transcripts and accumulated descriptions up to the paused frame. The generated content is reused from GenAD, reducing response latency and improving relevance. Initial tests suggest that, as with GenAD, strategic prompting yields responses that are more contextual and align with BLV users’ preference for concise narration.

\begin{figure}[H]
    \centering

    \begin{subfigure}[t]{0.85\linewidth}
        \centering
        \includegraphics[width=\linewidth]{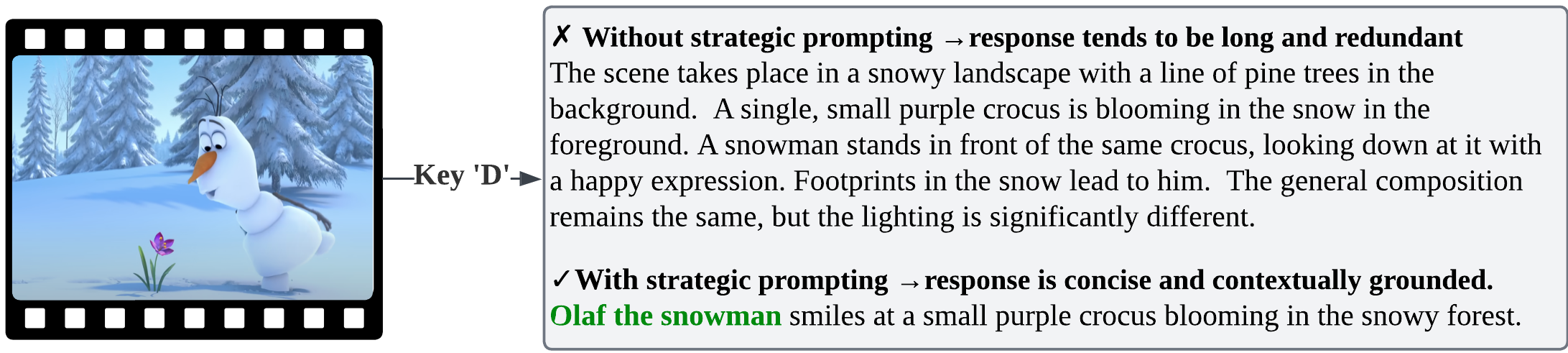}
        \caption{Frozen trailer – Without strategic prompting, the description is verbose and generic; with prompting, it becomes concise and specific to the context.}
        \label{olaf}
    \end{subfigure}

    \vspace{1em}

    \begin{subfigure}[t]{0.85\linewidth}
        \centering
        \includegraphics[width=\linewidth]{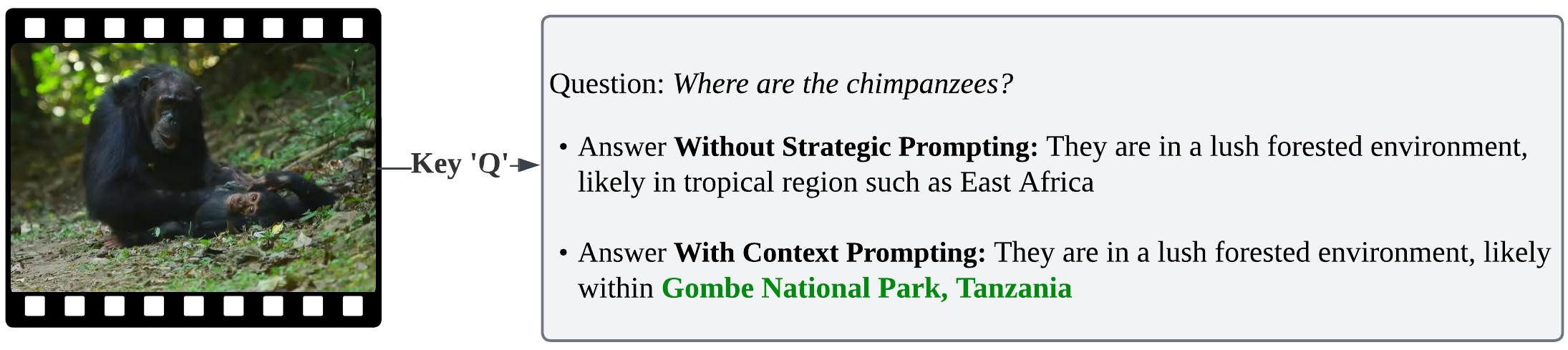}
        \caption{Gombe National Park video – Without context prompting, the model gives a vague regional guess; with context prompting, it correctly specifies Gombe National Park, Tanzania.}
        \label{chimpanzees}
    \end{subfigure}

    \caption{Examples of AdaptAD responses with and without accessibility-guided prompting.}
    \label{adaptad-examples}
\end{figure}

Fig.\ref{adaptad-examples} illustrates two cases. In the Frozen trailer (Fig.\ref{olaf}), we paused on a frame and pressed \textit{D} to request additional description. Without strategic prompting, the system returned a verbose and generic answer. With prompting, however, the output was brief and contextual, correctly naming the snowman: \textit{“Olaf the snowman.”} In the Jane Goodall-Gombe National Park video (Fig.~\ref{chimpanzees}), pressing \textit{Q} to ask \textit{“Where are the chimpanzees?”} initially yielded only a more general regional response. When supplemented with context from earlier scene descriptions, however, AdaptAD produced the precise answer: \textit{“Gombe National Park, Tanzania.”}

Prior work has already shown that interactive, on-demand descriptions are valuable for BLV users, allowing them to explore content in ways that fixed AD alone cannot. Building on this evidence, our work experiments with VLMs and strategic prompting to produce more contextual answers while integrating AdaptAD with the upgraded GenAD module. Although these preliminary observations are encouraging, further studies with BLV participants are needed to confirm how well the responses meet user needs and to identify opportunities for improvement.

\section{Formative Expert Assessment of GenAD}

To investigate the current strengths and weaknesses of VLMs for audio description, an evaluation workshop was conducted with seven accessibility consultants. These experts bring extensive experience working with BLV users in both professional and advocacy contexts, including teaching novice describers, designing accessible media resources, training BLV users on assistive technologies, and supporting community-based accessibility initiatives. Because factual accuracy was one of the evaluation dimensions, sighted experts were needed to verify whether descriptions faithfully represented on-screen content. While BLV users did not directly participate in this stage, accessibility consultants anchored the evaluation in BLV-centered principles and standards, while also verifying quality dimensions, such as accuracy, that require sighted review. As design research, the workshop offered formative insights into both the capabilities and shortcomings of GenAD.

\subsection{Video Selection}

\begin{figure}[H]
  \centering

\begin{subfigure}[t]{0.18\linewidth}
  \centering
  \includegraphics[width=\linewidth]{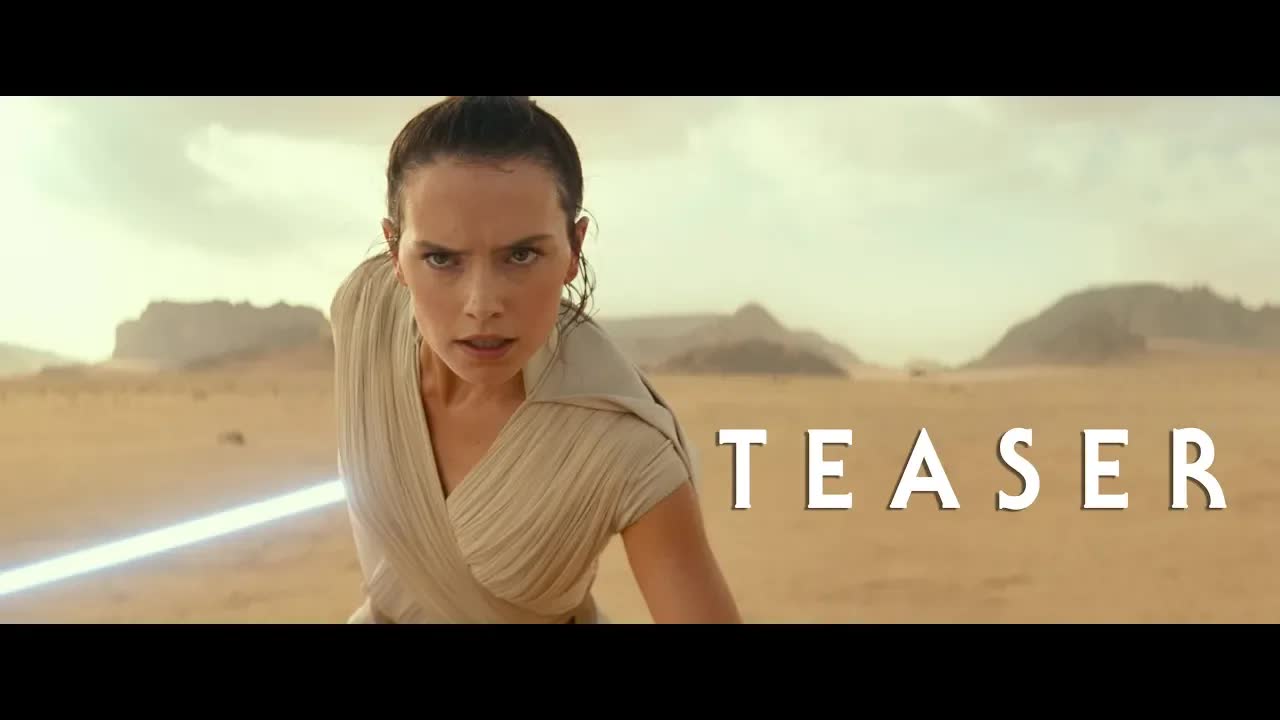}
  \caption{Star Wars: The Rise of Skywalker – Teaser}
\end{subfigure}
\hspace{0.015\linewidth}
\begin{subfigure}[t]{0.18\linewidth}
  \centering
  \includegraphics[width=\linewidth]{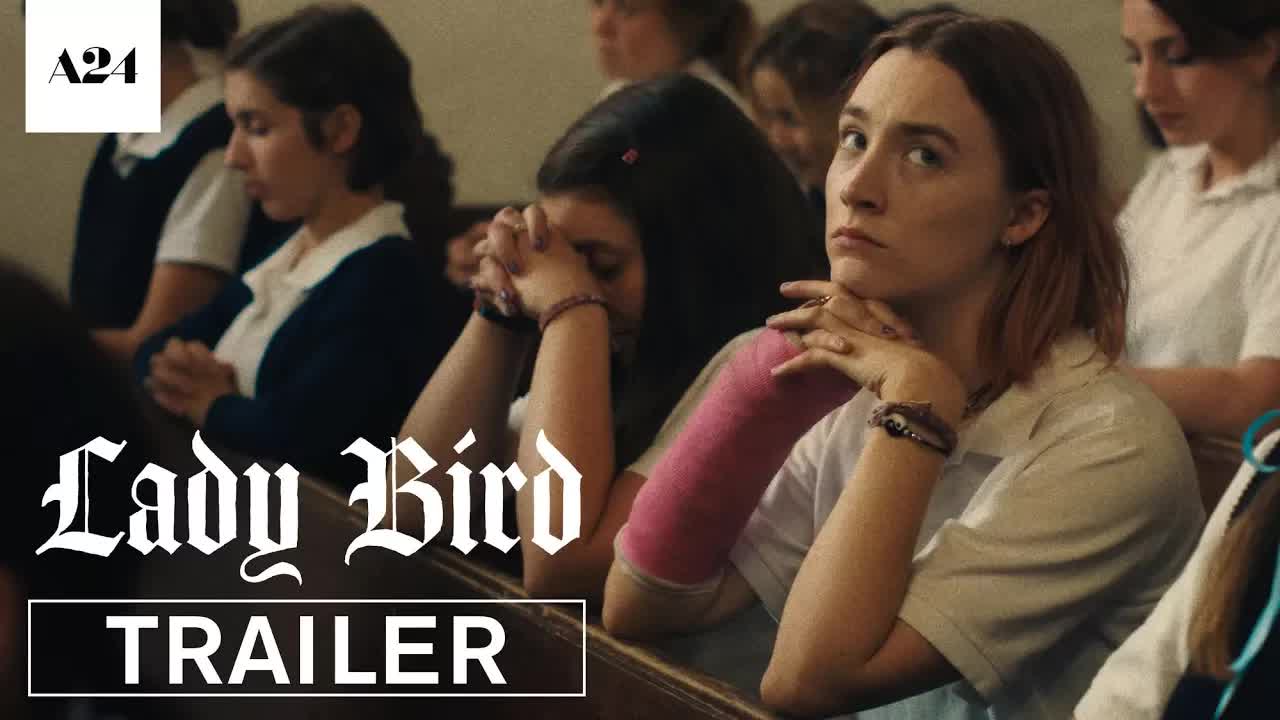}
  \caption{Lady Bird | Official Trailer HD | A24}
\end{subfigure}
\hspace{0.015\linewidth}
\begin{subfigure}[t]{0.18\linewidth}
  \centering
  \includegraphics[width=\linewidth]{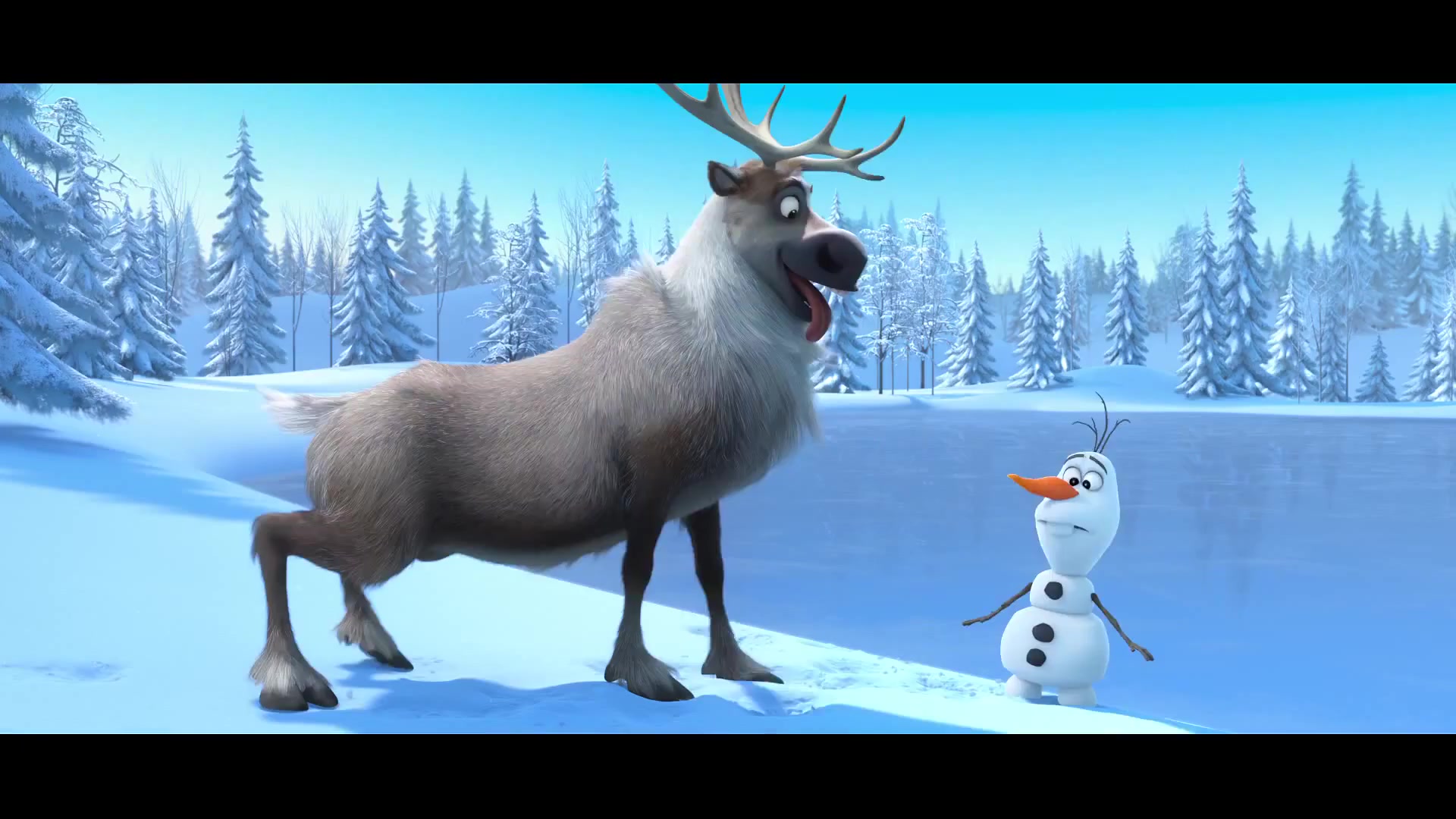}
  \caption{Frozen Teaser (2013) - Disney Animated Movie}
\end{subfigure}
\hspace{0.015\linewidth}
\begin{subfigure}[t]{0.18\linewidth}
  \centering
  \includegraphics[width=\linewidth]{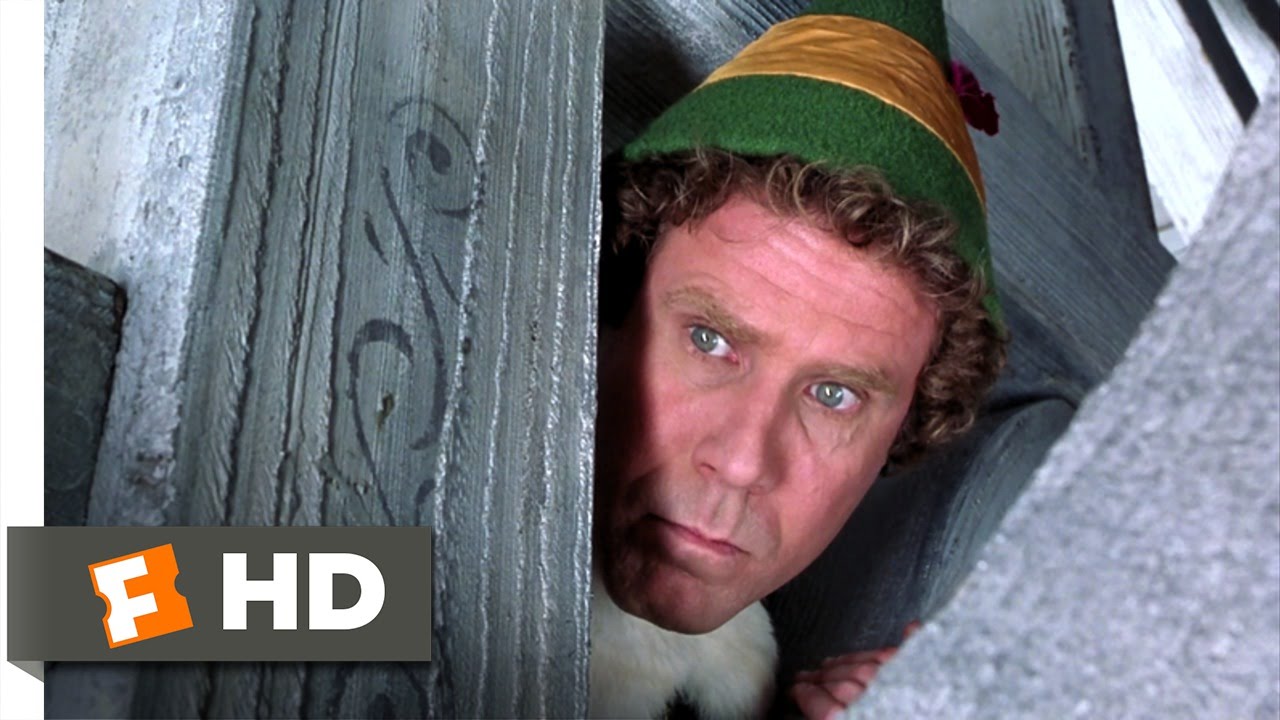}
  \caption{Elf Clip - Buddy Realizes He's Human (2003)}
\end{subfigure}
\hspace{0.015\linewidth}
\begin{subfigure}[t]{0.18\linewidth}
  \centering
  \includegraphics[width=\linewidth]{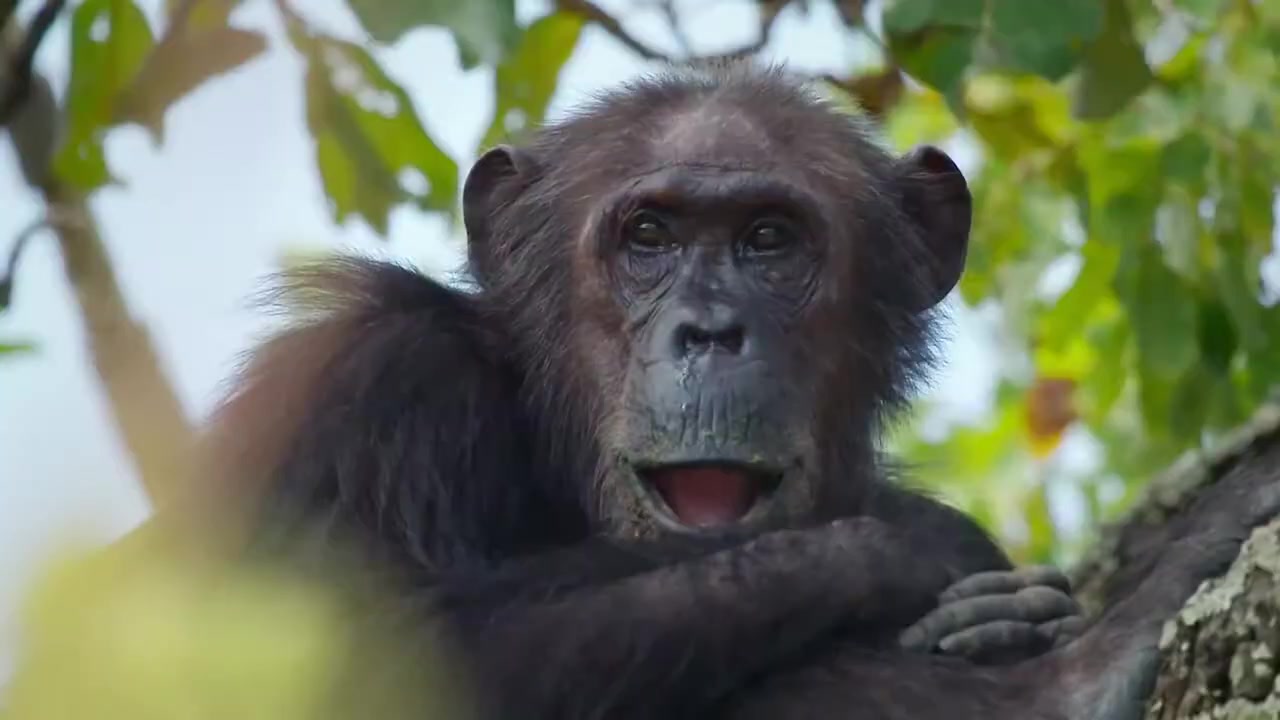}
  \caption{Jane Goodall - Gombe National Park}
\end{subfigure}

\begin{subfigure}[t]{0.18\linewidth}
  \centering
  \includegraphics[width=\linewidth]{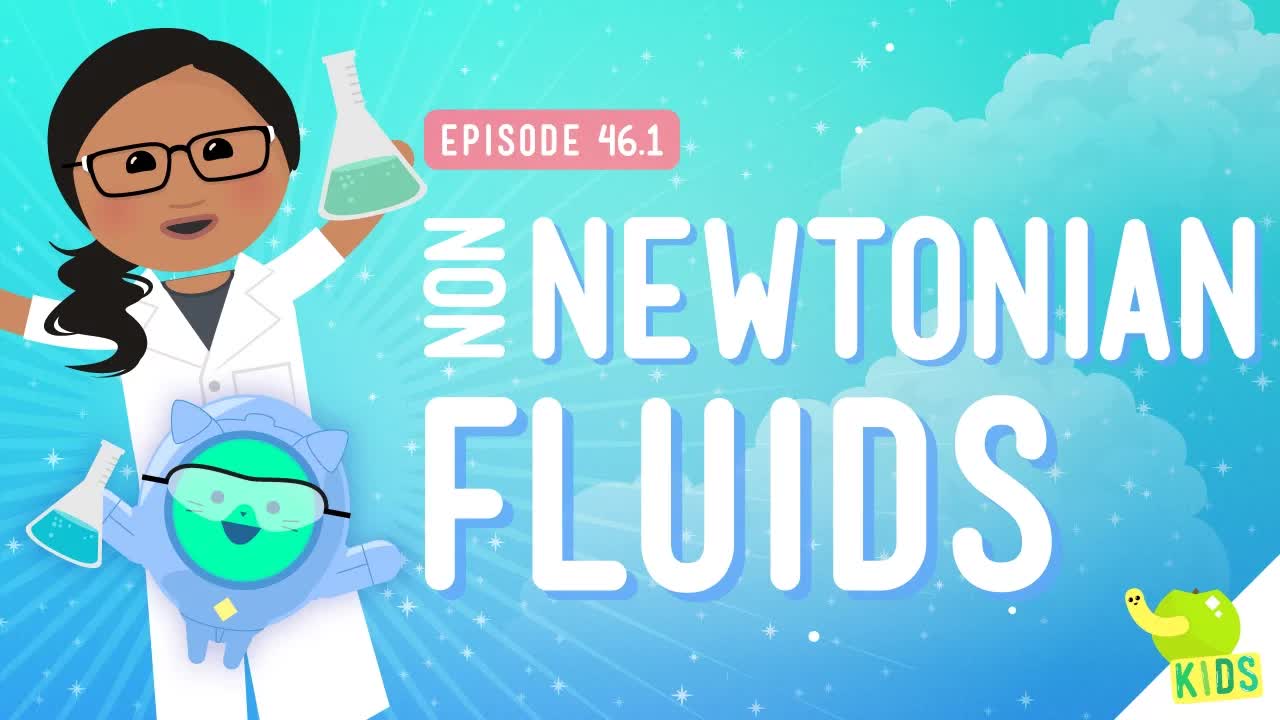}
  \caption{Non-Newtonian Fluids: Crash Course Kids}
\end{subfigure}
\hspace{0.015\linewidth}
\begin{subfigure}[t]{0.18\linewidth}
  \centering
  \includegraphics[width=\linewidth]{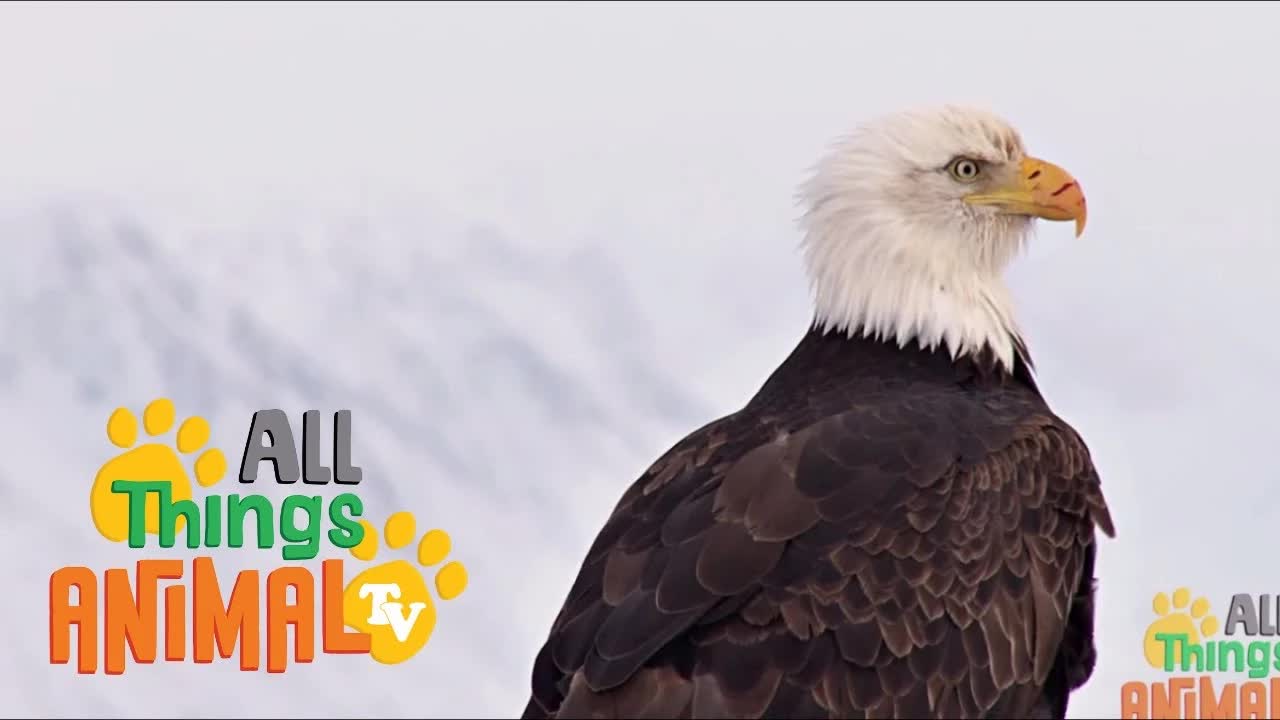}
  \caption{Bald Eagle | Animals for Kids | Animal TV}
\end{subfigure}
\hspace{0.015\linewidth}
\begin{subfigure}[t]{0.18\linewidth}
  \centering
  \includegraphics[width=\linewidth]{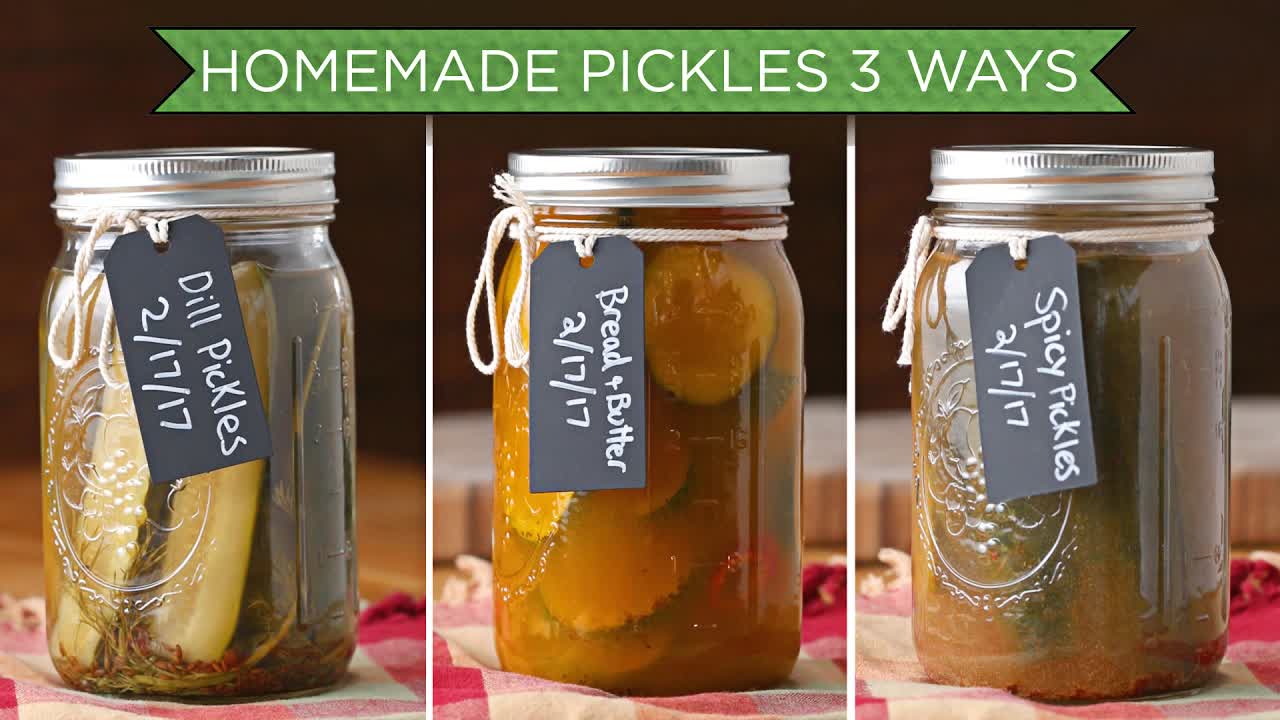}
  \caption{3 Ways to Make Homemade Pickles}
\end{subfigure}
\hspace{0.015\linewidth}
\begin{subfigure}[t]{0.18\linewidth}
  \centering
  \includegraphics[width=\linewidth]{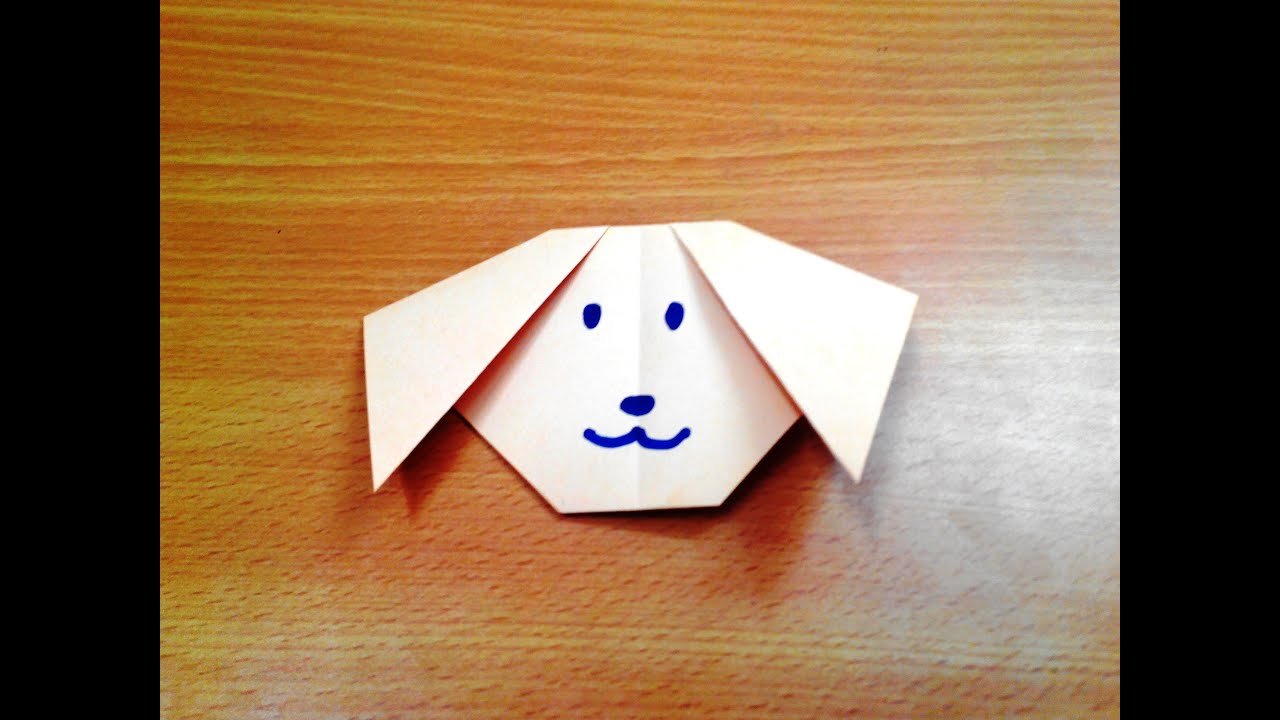}
  \caption{How to Make an Origami Dog Face}
\end{subfigure}
\hspace{0.015\linewidth}
\begin{subfigure}[t]{0.18\linewidth}
  \centering
  \includegraphics[width=\linewidth]{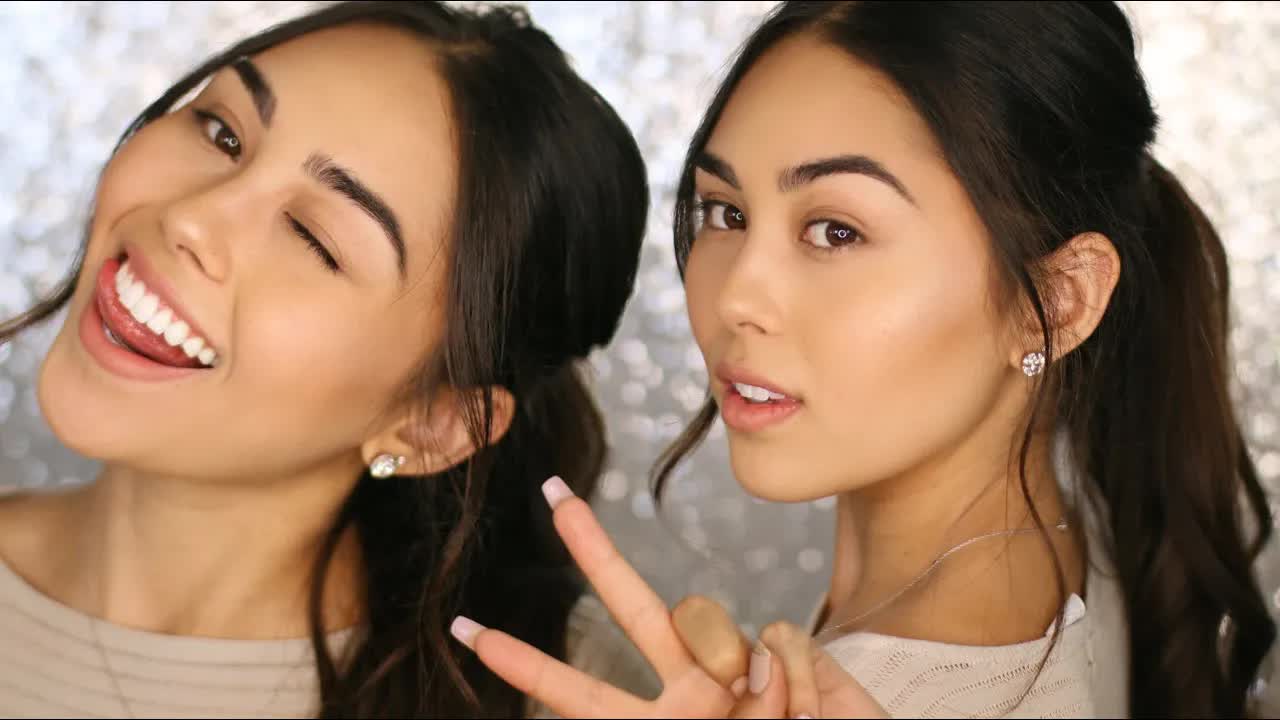}
  \caption{Quick, Easy 5-Minute Makeup Tutorial}
\end{subfigure}
  
  \vspace{1em} 

\caption{Ten YouTube videos used in the evaluation, spanning Entertainment, Education, and How-to \& Style. Subcaptions show the YouTube titles.}

  \label{fig:study-videos}
\end{figure}

Ten videos were selected across three categories, including Entertainment, Education, and How-to \& Style, to capture variation in genre conventions, audiences, and descriptive challenges.

Entertainment included four familiar clips: Star Wars, Lady Bird, Elf, and Frozen. These span action-driven sequences, teen drama, family comedy, and children’s animation. Their shifting tones, narrative density, and cinematic techniques such as pacing, visual effects, and atmosphere made them particularly demanding for nuanced description of both character interactions and mood.

Education featured a Crash Course explainer on non-Newtonian fluids, a children’s wildlife video on bald eagles, and a documentary excerpt with Jane Goodall at Gombe National Park. Each embodies a different teaching style: Crash Course compresses dense science into fast narration and colorful visuals; the Bald Eagle video simplifies concepts for children with engaging imagery; while the Jane Goodall excerpt reflects a more traditional documentary mode with conservation-focused narration. Descriptive challenges in this context centered on balancing factual precision with accessibility, while avoiding interference with the spoken dialogue, much of which was delivered through the video’s instructors and narrator.

How-to \& Style included three everyday instructional clips: a pickle recipe, a makeup tutorial, and an origami dog demonstration. Unlike narrative-driven content, these videos demanded precise description of small, sequential hand movements, tools, and materials. The rapid pace of short-form style made this particularly challenging, placing a premium on timing and clarity so BLV viewers could follow step-by-step processes without confusion or cognitive overload.

On average, the videos were about three minutes long (range: 1:30–5:04) and had been requested on community-driven AD platforms such as YouDescribe \cite{YouDescribeYouDescribe.Https://www.youdescribe.org/}. Their selection reflected genuine demand from BLV audiences rather than arbitrary choice. Collectively, the set provided a meaningful testing ground for evaluating how well VLMs adapt across genres, audiences, and visual modalities, spanning action-heavy trailers, comedic performances, task-oriented tutorials, animated science explainers, and observational documentary footage that BLV audiences actively seek out and request to be described.

\subsection{Workshop Design}

For each video, three AD versions were generated using an identical pipeline (Section 3.1) with Qwen2.5-VL, Gemini 1.5 Pro, and GPT-4o as the only variable. To minimize bias, anonymization and randomization were applied at three levels: (1) versions were relabeled A, B, and C with randomized mappings per video; (2) the order of the ten videos was randomized per consultant; and (3) within each video, the sequence of versions was randomized. These measures reduced ordering effects, prevented recognizable labeling, and encouraged independent evaluation.

Consultants viewed all three versions of a video sequentially, completing an assessment form immediately after each. This process was repeated for all ten videos, yielding 30 evaluations per consultant (210 total). While forms were submitted after each viewing, consultants could revisit and revise earlier responses during the session—supporting independent first impressions while allowing later reconsideration.

\subsubsection{Evaluation rubric} The evaluation rubric was grounded in the Described and Captioned Media Program (DCMP) guidelines, the most widely cited standards for educational and entertainment AD \cite{2024DescribedDCMP}. Five main content dimensions drawn from DCMP: Accurate, Prioritized, Appropriate, Consistent, and Equal. However, DCMP provides limited direction on delivery format. With crowdsourced platforms now supporting both inline and extended narration, and placement becoming an important part of the audio description process, we incorporated two delivery-focused dimensions, informed by NCAM broadcast guidelines \cite{NationalCenterforAccessibleMedia2017AccessibleGuidelines} and industry practices from 3Play Media \cite{3PlayMedia2020AudioGuidelines}.

The resulting rubric was organized into seven dimensions that together assess both content and delivery qualities of AD (see Appendix for the full documentation provided to consultants).

\begin{enumerate} 
    \item Accurate: Descriptions are factually correct and error-free. 
    \item Prioritized: Salient visual details are emphasized; trivial elements are omitted. 
    \item Appropriate: Language and level of detail match the genre and intended audience (e.g., clarity for education; engagement for entertainment). 
    \item Consistent: Terminology, style, and tone remain coherent throughout. 
    \item Equal: Descriptions use objective, unbiased language; avoids personal interpretation. 
    \item Strategic Use of Delivery Method (Inline/Extended) is chosen appropriately: Inline when natural pauses suffice; extended when critical details cannot otherwise fit. 
    \item Timing \& Placement: Descriptions align with visual events and avoid masking essential audio. \end{enumerate}

Consultants rated each dimension on a 1–5 scale (1 = major violations; 3 = adequate with minor issues; 5 = exemplary). The full rubric with behaviorally anchored examples is provided in the Appendix. This ensured judgments were standardized while capturing both textual accuracy and temporal integration with audiovisual material.

\subsection{Quantitative Results}

\begin{table}[h!]
\centering
\begin{tabular}{|l|c c|c c|c c|}
\hline
\multirow{2}{*}{\textbf{Criteria}} 
& \multicolumn{2}{c|}{\textbf{Qwen}} 
& \multicolumn{2}{c|}{\textbf{Gemini}} 
& \multicolumn{2}{c|}{\textbf{GPT}} \\
\cline{2-7}
& Mean & SD & Mean & SD & Mean & SD \\
\hline
\textbf{Overall}   & 3.78 & 1.00 & 4.01 & 1.02 & \textbf{4.05} & 0.97 \\
Accurate           & 3.63 & 1.13 & \textbf{3.94} & 1.08 & \textbf{3.94} & 1.13 \\
Prioritized        & 3.43 & 1.06 & 3.61 & 1.11 & \textbf{3.76} & 1.10 \\
Appropriate        & 3.90 & 1.06 & \textbf{4.16} & 1.08 & \textbf{4.16} & 1.06 \\
Consistent         & 4.03 & 0.97 & \textbf{4.31} & 1.02 & 4.27 & 0.96 \\
Equal              & 4.41 & 0.95 & \textbf{4.56} & 0.88 & 4.41 & 1.06 \\
Strategic Use of Delivery Method    & 3.37 & 1.01 & 3.67 & 1.15 & \textbf{3.84} & 0.98 \\
Track Placement    & 3.69 & 0.94 & 3.79 & 1.01 & \textbf{3.97} & 0.87 \\
\hline
\end{tabular}
\vspace{0.5em}

\caption{Overall and per-criteria AI scores with separate Mean and SD. Bold indicates the top-performing model per row.}
\label{tab:ai_scores}
\end{table}

\begin{figure}[h]
  \centering
  \includegraphics[width=\linewidth]{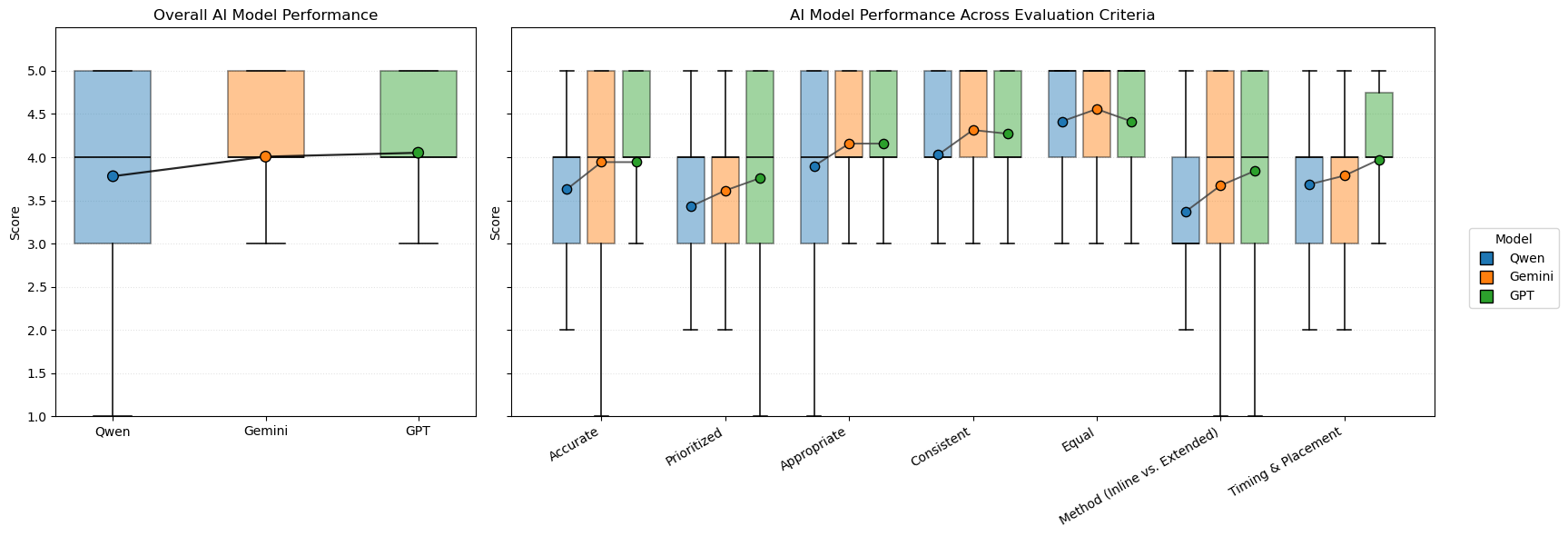}
  \caption{Distribution of rater scores for Qwen, Gemini, and GPT across the overall measure and seven criteria. Boxplots show score variability; dots mark means connected within each category}
  \label{quantitative}
\end{figure}

\subsubsection{Overall Mean Scores}

As shown in Table \ref{tab:ai_scores}, the mean scores across models were relatively close, with GPT-4o at 4.05, Gemini 1.5 Pro at 4.01, and Qwen2.5-VL at 3.78. While the absolute differences are modest, the distributions presented in Fig. \ref{quantitative} reveal differing variance patterns across models. Qwen2.5-VL shows scores across the full 1-5 range, while GPT-4o and Gemini 1.5 Pro demonstrated more concentrated distributions, with the majority of scores falling within 3-5 range. These distribution patterns indicate that Qwen produces more variable output quality, while GPT and Gemini show greater consistency in their performance scores. 

\subsubsection{Dimension-Level Performance}

The dimension-level scores (Table~\ref{tab:ai_scores}, Fig.~\ref{quantitative}) show that Consistent, Equal, and Appropriate had averages between 4.0 and 4.3. Ratings for these dimensions were primarily in the 4–5 range, with relatively few evaluations below 4. This pattern suggests that across videos, evaluators generally rated the models as maintaining stable terminology, neutrality, and alignment with genre style. The Accurate dimension received moderate scores overall (means 3.6–3.9), but with distinct distribution patterns across models. Qwen had the lower mean (3.63), with most ratings between 3 and 5 and a few dipping to 2. Gemini and GPT shared the higher mean (3.94), but their distributions differed; GPT’s ratings clustered around 4, while Gemini spanned the full 1–5 scale, indicating greater variability. The Prioritized dimension had the lowest mean scores among the evaluated dimensions including both content and formatting, with averages between 3.3 and 3.7. GPT’s ratings in this area covered a broader range, while Qwen and Gemini were more frequently rated between 3 and 4 and never peaked at the top score. Since prioritization is inherently subjective, depending on what evaluators judge as relevant in different types of content, scores tended to be lower overall and more variable.

The two formatting dimensions, Strategic Usage of Delivery Method (Inline/Extended) and Timing \& Track Placement, address how descriptions are integrated into the audiovisual stream rather than their content. Their averages were generally lower than those of the content dimensions Accurate, Consistent, Equal, and Appropriate, with the exception of Prioritized. Strategic Usage of Delivery Method narration scored between 3.4 and 3.8, with ratings spanning a wide range. This variability likely reflects differences in how evaluators judged the necessity of extended narration: while inline delivery is typically preferred, opinions diverged on when important visual details justified inserting extended descriptions during playback. Timing \& Track Placement scored between 3.7 and 4.0, with rating distributions showing a breadth similar to Strategic Use of Delivery Method. This may reflect variation in how descriptions were aligned with video timing and dialogue.

Overall, there was a distinction between content-related and formatting-related dimensions. Consistent, Equal, and Appropriate showed relatively narrow spreads, with most ratings clustered toward the higher end of the scale. Accurate had somewhat lower scores with occasional dips, while Prioritized received the lowest scores across all dimensions and displayed wider distributions. The two formatting dimensions were also rated below most of the content dimensions (except Prioritized) and showed broader spreads, indicating greater variability in evaluations of delivery choices and synchronization with the audiovisual stream. While these findings should be interpreted with caution given the small sample size, they highlight opportunities to refine GenAD’s performance, particularly in producing prioritized and well-formatted descriptions to deliver more meaningful audio description for BLV users.

\subsubsection{Video Category-level Performance}

\begin{table}[h!]
\centering
\begin{tabular}{|l|c|c|c|}
\hline
\textbf{Video Category} & \textbf{Qwen} & \textbf{Gemini} & \textbf{GPT} \\
\hline

\textbf{Entertainment} (Star Wars, Lady Bird, Elf, Frozen) & 3.66 & \textbf{3.91} & 3.87 \\
\hline

\textbf{Educational} (Crash Course Kids, Bald Eagle) & 4.01 & 4.14 & \textbf{4.27} \\
\hline

\textbf{How-to} (Pickles, Origami, Makeup) & 3.76 & 4.12 & \textbf{4.14} \\
\hline
\end{tabular}
\vspace{0.5em}
\caption{Mean AI scores across video categories. Bold indicates the best model per row.}
\label{tab:video_categories}
\end{table}

Performance varied across genres (Table~\ref{tab:video_categories}). Across all three models, Educational clips received the highest mean scores, followed by How-to \& Style, with Entertainment rated lowest. This ordering (Educational > How-to > Entertainment) was consistent regardless of model. GPT and Gemini scored above Qwen in every category, with GPT slightly ahead in Educational and How-to, and Gemini slightly ahead in Entertainment.

In summary, these results show that the relative ranking of models was stable across genres, with GPT and Gemini generally producing higher averages and Qwen consistently lower. They also signal that genre played a role in overall scoring, with fact-oriented Educational content evaluated more favorably than task-oriented How-to or narrative-driven Entertainment clips. The lower scores for Entertainment may reflect the added complexity of describing cinematic material, where dense visual detail and stylistic elements can make it harder to decide which aspects to include in ways that remain both clear and engaging for viewers. These patterns should be read as exploratory observations only, given the limited number of videos in each category and the small pool of evaluators.

\subsection{Qualitative Discussion}

 Feedback from consultants highlighted both the strengths and the limitations of AI-generated audio description. Several raters described some tracks as \textit{“very good”} or \textit{“lovely,”}  with one even remarking they were \textit{“curious if it is human or AI.”} One consultant, who trains both sighted volunteers and BLV users in accessibility tools, observed that \textit{“AI has come a long way”} and that some outputs were \textit{“comparable”} to those of novice describers. At the same time, consultants emphasized that the descriptions still lacked \textit{“that last 10\%”} needed to make them fully enjoyable for BLV audiences, and in some cases \textit{“simply do not compare”} to the professional AD content typically encountered on broadcast media and streaming platforms. Several noted that AI tends to make similar judgment errors as beginners, such as over-describing visual details, speaking over important music or sound effects, using excessive signposting or transitions, and relying on repetitive or generic word choices, highlighting the need for editorial oversight and refinement.

One of the most frequently cited weaknesses concerned timing and placement. Raters described instances where descriptions were delivered \textit{“way too early,”} interrupted dialogue mid-word in ways they found \textit{“irritating,”} or were compressed into bursts that were \textit{“fast and hard to follow.”} For example, in movie trailers, poor temporal alignment stemmed from AI attempting to pack dense visual information into limited pauses, often clashing with dialogue and speech rhythm. Another example in how-to videos, even without dialogue, their inherently rapid pace and emphasis on ease made rushed descriptions especially difficult to follow. One evaluator noted that this issue could have been alleviated through the use of extended narration, which would have provided more space to keep up.

However, heavy reliance on extended narration was not always effective, since prioritization also emerged as a recurrent problem. While AI descriptions were often accurate, raters pointed to an excess of \textit{“superfluous description”} and a tendency to describe information already inferable from dialogue. Because prioritization is inherently subjective, what one viewer considers important may not matter to another, consultants suggested that user-driven tools could help address this gap by giving BLV users agency to request additional details on demand.

Several evaluators observed that AI-generated AD often fell into the space of \textit{“good, but not elegant.”} One consultant summarized this sentiment by noting, \textit{“I think with just a few tweaks this could go from very good to excellent.”} This theme points to the value of a human-in-the-loop (HITL) workflow: with human describers refining drafts, small adjustments in prioritization and formatting could transform adequate descriptions into ones that feel natural and seamless.

Notably, the evaluators drew a distinction between adequacy for everyday consumption and adequacy for instructional use. Several AI-generated outputs were considered \textit{“good enough to watch”} by BLV audiences, yet fell short of the standards required for teaching novice describers. One rater remarked, \textit{“I would never use this video as a teaching tool; I’d create my own.”} This points to a divergence in evaluative standards: consultants adopting a consumer lens were more forgiving of imperfections, while those viewing through a pedagogical lens demanded the precision, restraint, and temporal elegance expected of professional AD.

Overall, raters were impressed with the fluency and informativeness of the outputs, but they consistently identified weaknesses in prioritization, use of inline versus extended narration, and timing and placement, patterns that mirror the quantitative results. These gaps, such as over-description, poor alignment with dialogue, or misplaced emphasis, were most disruptive in dialogue-heavy or fast-paced videos. The findings point to opportunities for pairing automation with refinement and interactivity: HITL workflows to strengthen editorial decisions and delivery, and AdaptAD to give BLV users greater control over which details they wish to prioritize.

\section{RefineAD in Practice} 

As one expert remarked, \textit{“with a few tweaks this can go from good to excellent,”} we invited them to experiment with the HITL workflow by using the editing interface (Fig. \ref{colab}) to refine AI-generated descriptions. Due to time and budget limitations, we could not ask all consultants to complete the task across all videos. Instead, the expert selected to edit one AI-generated version for four videos: Frozen Teaser Trailer (2013), Jane Goodall – Gombe National Park, 3 Ways to Make Homemade Pickles and How to Make an Origami Dog Face . These selections were arbitrary and not based on video type, AI system, or initial quality. The consulant emphasized that while editing still required some effort, the workflow was both quicker and easier than starting from scratch—advantages that would likely be even more significant for novice describers.

\begin{figure}[H]
  \centering
  \includegraphics[width=\linewidth]{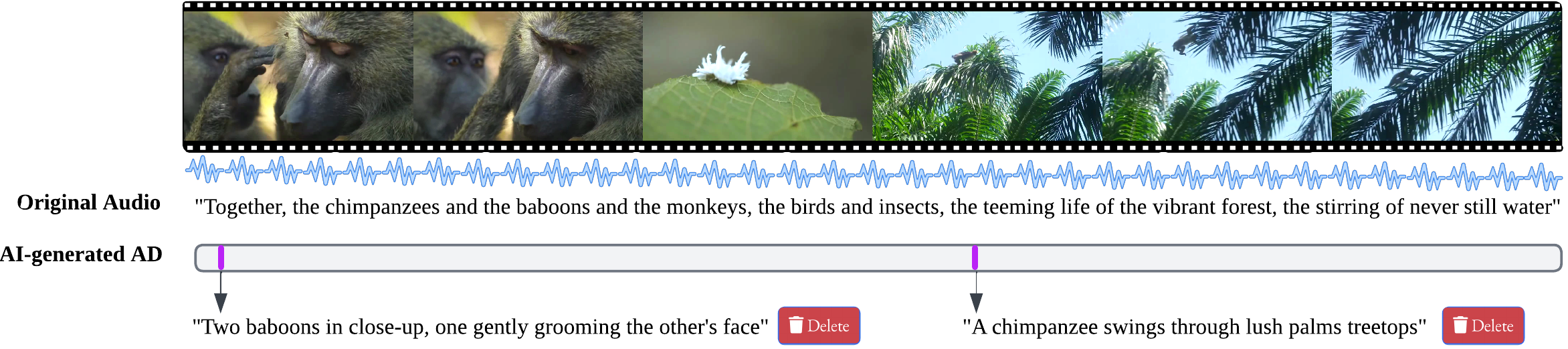}
  \caption{Example from the Jane Goodall video. Extended descriptions generated for visual details interrupted the flow of Jane Goodall’s narration. Through the editing interface, such tracks can be deleted to maintain the natural rhythm of the original audio.}
  \label{jane-hitl}
\end{figure}

The Jane Goodall video illustrated how description delivery remain a challenge. As shown in Fig.~\ref{jane-hitl}, the spoken narration is already evocative—painting a vivid image of \textit{“chimpanzees and baboons and monkeys, the birds and insects, the teeming life of the vibrant forest, the stirring of the never still water.”} In this context, the extended descriptions disrupt the immersive quality of her narration by pausing the video to insert details such as \textit{“two baboons in close-up...”} or \textit{“a chimpanzee swings through lush palm treetops.”} Although accurate, the expert observed that these insertions repeated information already suggested by the spoken track and ultimately detracted from its rhythm. Through the collaborative editing interface, the editor could easily delete such tracks, resulting in a tighter version that better preserved the flow of the original audio.

\begin{figure}[H]
  \centering
  \includegraphics[width=\linewidth]{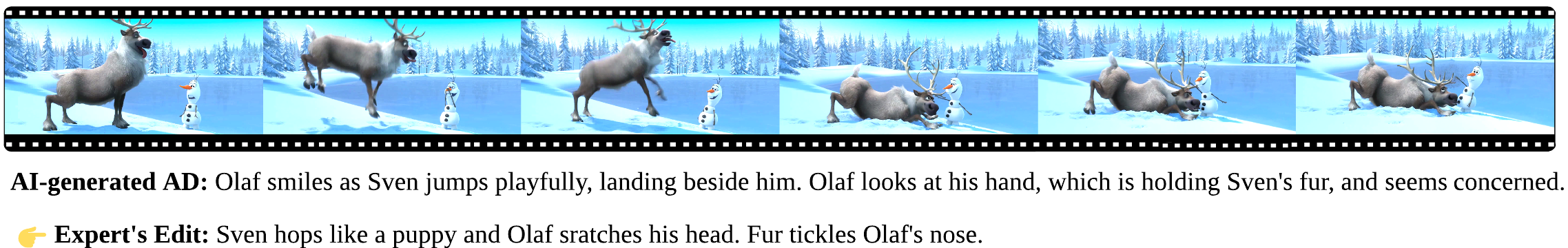}
  \caption{Example from the Frozen Teaser. The expert’s edit transforms a literal AI draft by adding playful comparisons and tighter phrasing, producing a narration that is more engaging and age-appropriate.}
  \label{frozen-hitl}
\end{figure}

A clear example of how the HITL component can provide the \textit{“last 10\%”} of improvement noted in expert feedback comes from the Frozen trailer (see Fig. \ref{frozen-hitl}). In this scene, the AI-generated description is factually correct in noting that Sven jumps beside Olaf. However, the expert’s edit adds nuance by choosing the verb “hops” and comparing Sven to a puppy, while also noting how Olaf “scratches his head.” These choices not only depict the action but also highlight its playful, childlike quality—well aligned with the tone of an animated film for children. Similarly, the expert revised “Olaf looks at his twig arm which is holding Sven’s fur and looks surprised” to the much shorter “Fur tickles Olaf’s nose.” The edit conveys the same visual information more vividly and efficiently, reinforcing humor and character while avoiding unnecessary detail. 

In How-to videos, timing and prioritized details are crucial—two areas where the expert noted AI-generated descriptions often fell short. Several lines required placement adjustments. As the expert observed, some cues were delivered \textit{“too early,”} and when actions were described \textit{“too fast,”} BLV audiences would struggle to replicate the steps. These issues were easily corrected by adjusting start times or using the \textbf{nudging tool} in the collaborative editing interface (Fig.~\ref{colab}). Aligning narration with on-screen events, rather than anticipating them, provided clearer pacing that gave BLV audiences enough time to follow each step and replicate the actions.

\begin{figure}[H]
  \centering
  \includegraphics[width=\linewidth]{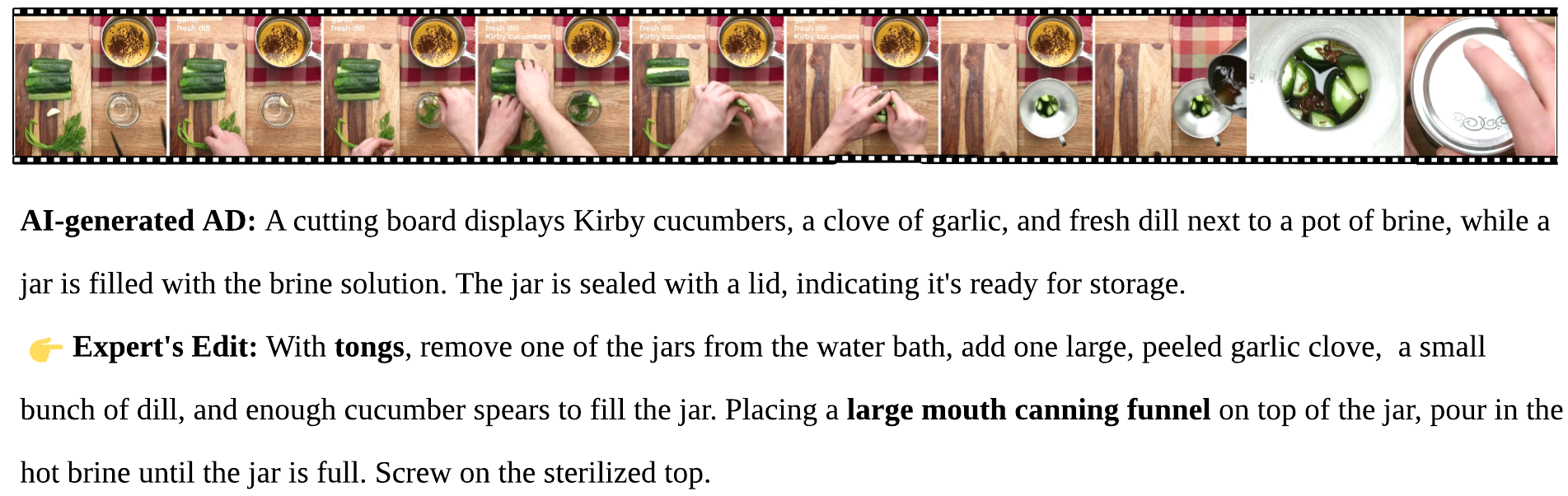}
  \caption{Example from the 3 Ways to Make Homemade Pickles clip. Expert edits expanded on AI-generated actions by naming tools and clarifying procedure, making the steps more practical and replicable.}
  \label{pickles-hitl}
\end{figure}

A clear example of how expert edits add essential details can be seen in the 3 Ways to Make Homemade Pickles clip (Fig.~\ref{pickles-hitl}). The AI-generated descriptions identified the main ingredients and outlined basic actions, but in a fairly general manner. The expert expanded these into precise procedural steps, explicitly naming tools, such as using \textit{“tongs”} to lift jars or a \textit{“funnel”} to pour brine, and arranging actions in a logical sequence. This shift turned a broad overview into actionable, step-by-step instructions that the audiences could realistically follow. By emphasizing tool use, the edit also highlighted tactile elements—like handling hot jars or aligning a funnel—that are critical for replication, especially for BLV users, yet often overlooked in automated drafts.

\begin{figure}[h]
  \centering
  \includegraphics[width=\linewidth]{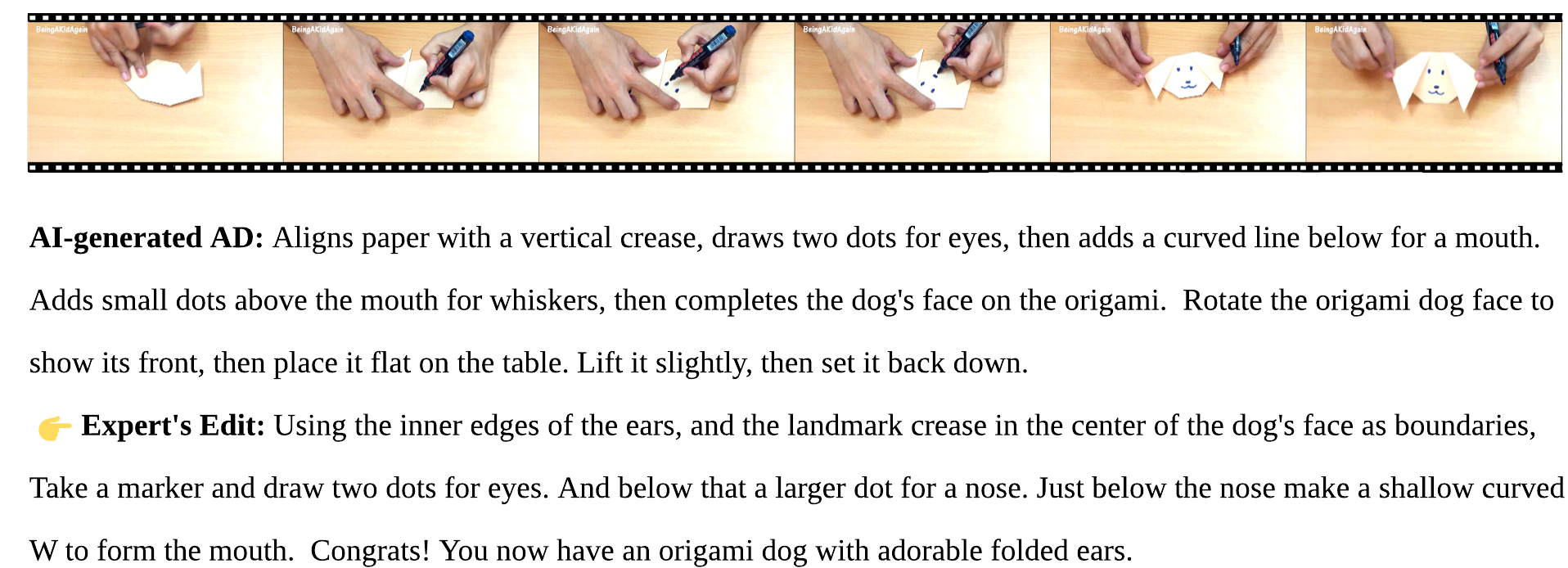}
  \caption{Example from the Origami clip. Expert edits emphasized spatial orientation and added a conversational closing, improving clarity and engagement}
  \label{origami-hitl}
\end{figure}

In the How to Make an Origami Dog face video (Fig.~\ref{origami-hitl}), the AI-generated narration accurately captured the basic steps for drawing the dog’s face, but the expert enriched these instructions with specific spatial references. For example, the expert highlighted the \textit{“inner edges of the ears”} and the \textit{“central crease of the dog’s face,”} providing clear landmarks to guide viewers. The whiskers were also described as a \textit{“curve W”} shape, a concise visual cue that made the instruction clear and engaging. Toward the end, the AI described the instructor’s actions literally as \textit{“place it on the table, lift it slightly, then set it back down.”} While correct, this phrasing can be confusing for audiences, since these gestures simply showcased the finished product rather than introducing new steps. The expert instead added, \textit{“Congrats! You now have an origami dog with adorable folded ears.”} This alternative conveyed completion and gave the narration a warm, natural tone that the AI draft lacked.

Together, the edits illustrate how HITL workflows can transform serviceable but uneven drafts into outputs that are leaner, clearer, and better aligned with the audiovisual stream. By trimming redundancies, clarifying procedural steps, and correcting timing, the consultant demonstrated that even modest interventions can significantly elevate the utility and fluidity of AI-generated descriptions.

\section{AdaptAD for User-Driven Experience}

There is no one-size-fits-all approach to audio description. While \textit{“good”} AD can be assessed against established guidelines, as one expert noted, \textit{“the ultimate judges of the audio description are blind and low vision users who consume the content.”} Preferences vary not only across individuals but also across video genres and contexts, shaped by personal experience and viewing goals \cite{JiangCrescentiaJungMahikaPhutaneItsScenarios}. This was evident in both our quantitative and qualitative workshop discussions: GenAD frequently struggled with prioritization and delivery, particularly when deciding between inline and extended narration. Because extended narration is conventionally defined as “used only when necessary,” the key challenge becomes deciding what counts as essential, a determination that is highly subjective. Some BLV viewers prefer concise tracks that minimize redundancy, while others want elaboration to satisfy curiosity or enable deeper engagement.

AdaptAD was designed to address this challenge by shifting control to the user. Instead of relying on a single editorial judgment, AdaptAD supports BLV viewers to request additional details on demand. Preliminary demonstrations across three examples illustrate how this functionality can enhance accessibility by filling gaps left by traditional AD systems. Importantly, these scenarios were shaped by consultants with experience working directly with BLV users, who shared examples of the types of questions audiences often raise.

\begin{figure}[h]
    \centering

    \begin{subfigure}[t]{0.55\linewidth}
        \centering
        \includegraphics[width=\linewidth]{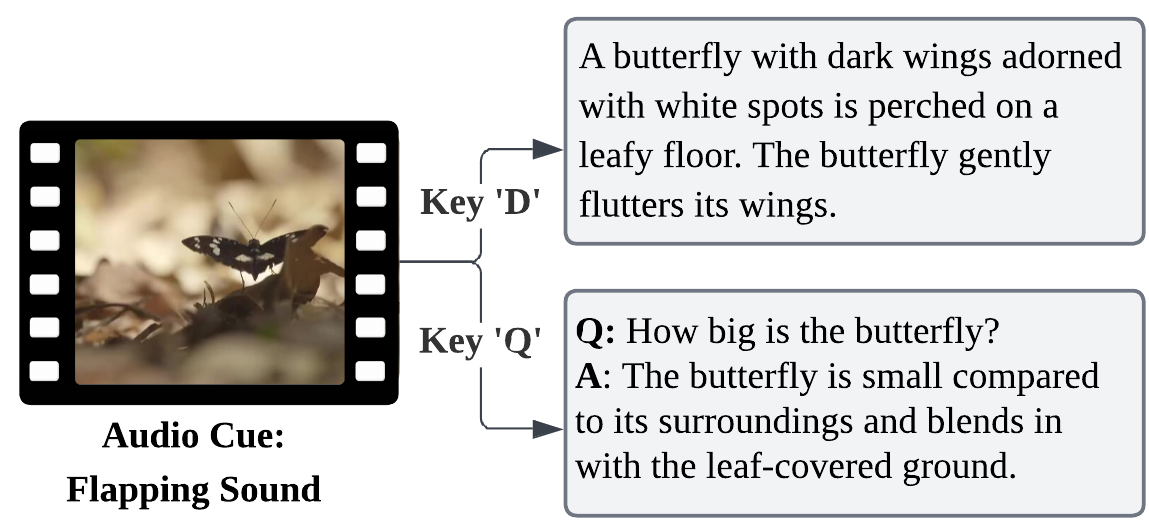}
        \caption{Butterfly scene in Jane Goodall – Gombe National Park (Education)}
        \label{jane-infobot}
    \end{subfigure}

    \vspace{1em}

    \begin{subfigure}[t]{0.55\linewidth}
        \centering
        \includegraphics[width=\linewidth]{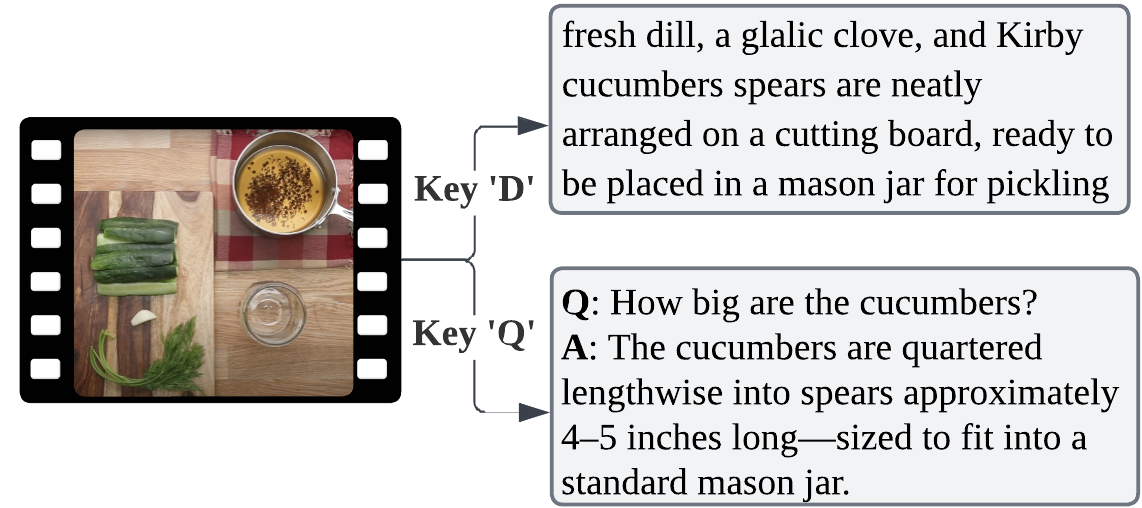}
        \caption{3 Ways to Make Homemade Pickles (How-to \& Style)}
        \label{pickles-infobot}
    \end{subfigure}

    \vspace{1em}

    \begin{subfigure}[t]{0.55\linewidth}
        \centering
        \includegraphics[width=\linewidth]{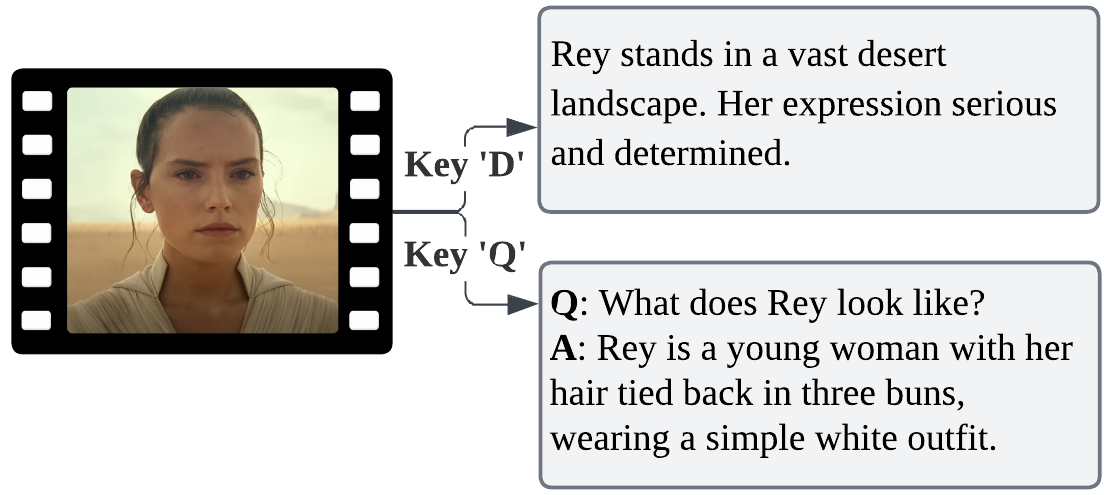}
        \caption{Star Wars: The Rise of Skywalker teaser (Entertainment)}
        \label{rey-infobot}
    \end{subfigure}

    \caption{Examples of AdaptAD responses triggered by audio cues or user queries, providing additional scene details on demand.}
    \label{fig:AdaptAD-trio}
\end{figure}

In the butterfly scene from Jane Goodall – Gombe National Park (Fig.~\ref{jane-infobot}), Jane Goodall’s narration already provides a rich account of the setting, so some audiences may feel an extended description of the butterfly is unnecessary and even disruptive. Others, however, may notice the flapping sound in the audio and wonder whether it comes from a bird or a butterfly. AdaptAD supports each user to decide: by pressing \textit{D}, they receive a description such as \textit{“A butterfly flutters across the frame.”} Those who remain curious can press \textit{Q} to ask a follow-up like \textit{“What kind of butterfly is it?”} and receive a detailed answer. In this way, AdaptAD accommodates both preferences without forcing a single editorial choice on all listeners.

Instructional videos pose similar challenge. In 3 Ways to Make Homemade Pickles (Fig.~\ref{pickles-infobot}), some viewers may prefer concise narration that avoids overloading specifics, while others require precision to replicate the recipe. Pressing \textit{D} provides an overview of the scene, for example:\textit{ “fresh dill, garlic cloves, and Kirby cucumber spears...”} For finer detail, a user can press \textit{Q} and ask: \textit{“How big are the cucumber pieces?”} to receive the answer: \textit{“They are quartered lengthwise into spears approximately 4–5 inches long.”} AdaptAD ensures that users seeking greater instructional clarity can access it, while those looking for a more efficient viewing experience are not burdened by unnecessary detail.

Entertainment contexts further illustrate the high degree of subjectivity in AD preferences. In the Star Wars: The Rise of Skywalker teaser (Fig.~\ref{rey-infobot}), some viewers may want additional detail about world-building and settings, while others are more interested in characters or even the actors portraying them. Consultants noted that BLV users frequently ask about an actor’s appearance and style. For example, a viewer might press \textit{Q} to ask: “\textit{What does Rey look like?”} AdaptAD could respond with a factual description such as: \textit{“Rey is a young woman with her hair tied back in three buns, wearing a simple white outfit.”} This allows users with a greater interest in character appearance to access that information on demand.

These examples illustrate why prioritization in AD cannot be solved by a single editorial choice: what feels superfluous to one viewer may be essential to another. By giving BLV audiences the ability to request quick clarification or deeper elaborations, AdaptAD makes audio description more flexible, personalized, and responsive to individual needs. Consultants also noted that the challenge is not only about balancing preferences, but about filling gaps that fixed narration can leave unaddressed. For instance, one consultant recalled working on a long, dialogue-free Baby Gorilla Cam video where describers failed to explain what gorillas actually look like until very late in the program. A blind quality controller caught this oversight, prompting the team to revise the description. Situations like this highlight how sighted describers, who are often “retinal-dependent,” may overlook what seems obvious to them but essential to BLV audiences. With AdaptAD, these gaps can be softened or eliminated, since users could simply ask questions such as \textit{“What does a gorilla look like?”} or \textit{“How does a gorilla compare to a human?”}.

The design of AdaptAD was guided by input from accessibility consultants with both lived experience and professional expertise, as well as by insights from prior studies involving BLV participants \cite{Bodi2021AutomatedUsers, Stangl2023TheVideos}. While these demonstrations are promising, the top priority moving forward is to evaluate the integrated system with BLV users, focusing on how often they use the \textit{D} and \textit{Q} functions, what prompts their use, and the types of questions they ask. Such data will help surface persistent gaps, guide refinements, and reveal what information BLV audiences most value.

Echoing what we learned from our study with accessibility experts, AdaptAD is designed not only as an integral part of the AD workflow but also as a mechanism for users to actively shape the process. Rather than passively consuming descriptions, BLV viewers can signal where narration feels incomplete and what details are most meaningful. Looking ahead, this functionality could also empower BLV individuals as content creators, enabling them to record and publish their own versions of audio description, tailored both to what they find important and to what they anticipate other BLV users value. In this way, AdaptAD has the potential to make content not only more accessible, but also more inclusive, responsive, and community-driven.

\section{Discussions}

\subsection{Reflection on Research Questions}
Our study set out to answer two research questions: (1) How well do modern VLMs generate baseline descriptions? and (2) How does the integration of GenAD, RefineAD, and AdaptAD improve narration quality and user experience?

For \textbf{RQ1}, our preliminary analyses showed that incorporating accessibility-guided prompting and contextual cues (e.g., audio transcripts, prior scene descriptions, video metadata) produced descriptions that were more relevant, cohesive, and aligned with BLV users' preferences. Expert evaluation suggested that VLMs such as Qwen, Gemini, and GPT can produce adequate drafts, with higher ratings on appropriateness, consistency, and equality, alongside moderate score on accuracy. However, weaknesses were evident in prioritization, delivery decision (inline vs. extended), and timing, which are central to AD practice. These results indicate that while VLMs can produce useful starting points, more design efforts are needed for a fully accessible system. 

For \textbf{RQ2}, prior studies have demonstrated the value of editing and interactive tools, even with earlier image captioning models \cite{Yuksel2020Human-in-the-LoopUsers, Bodi2021AutomatedUsers, Stangl2023TheVideos}. Our workshop with accessibility experts and formative demonstrations highlight how RefineAD and AdaptAD can precisely target GenAD’s weaknesses. RefineAD brings the human judgment needed for prioritization, delivery method, and track placement, while AdaptAD empowers BLV users to request the details most relevant to them, directly filling gaps that fixed narration cannot.

\subsection{Limitations and Future Works}

This study has several limitations. While the evaluation workshop with accessibility experts yielded valuable insights, quantitative results should be interpreted cautiously. The video set was narrow, covering only three genres, so findings remain descriptive and cannot support strong statistical conclusions. It is not yet possible to determine which of the three VLMs tested, Gemini, GPT, or Qwen, is most production-ready, nor to claim generalizable performance differences. A larger and more diverse dataset, combined with additional evaluators, will be necessary for statistically meaningful benchmarking.

Evaluation also presents challenges. Obtaining expert feedback at scale is difficult because assessing complete video descriptions is labor-intensive, requires attention to linguistic and temporal detail, and is costly. To address this, we will explore hybrid approaches that combine expert judgment with automated assessment. Recent studies suggest that LLMs can serve as evaluators, often aligning with human ratings \cite{ZhengJudgingArena, KocmiLargeQuality}. We aim to align such automated methods with accessibility experts’ standards to achieve reliable and scalable assessments. Rigorous quality control will be essential for AI-generated evaluations to be considered trustworthy.

A more significant limitation is the absence of direct studies with BLV participants. Although RefineAD was designed with accessibility in mind, incorporating WAI-ARIA for screen reader compatibility, full keyboard navigation, and other inclusive features, a complete, systematic evaluation with BLV audiences has not yet been conducted. Prior work highlighted that HITL editing and on-demand interactive AD can improve system quality and user experience \cite{Yuksel2020Human-in-the-LoopUsers, Stangl2023TheVideos}. Building on this foundation, our study upgraded GenAD with the expectation that improved baselines would reduce editing effort and increase satisfaction. Ultimately, however, BLV viewers are the gold standard for judging AD quality. We plan to conduct both controlled usability studies and longitudinal deployments to examine how BLV audiences engage with the system, and how their edits and preferences compare with those of sighted describers.

Addressing these limitations opens several promising directions. Broader datasets with greater variability across genres will be critical for testing GenAD against both professional and community-authored AD. Expanding evaluation to include a wider range of respondents including experts, professional describers, novice describers, and most importantly BLV participants, will ensure that assessment reflects the perspectives of those who rely on AD most.

Our findings suggest a need to revisit evaluation frameworks. Overall, consultants found the assessment model clear, easy to use, and aligned with professional guidelines. Some noted that delivery-focused dimensions, such as inline vs. extended narration and track placement, may warrant greater emphasis than stylistic qualities like consistency. Defining and measuring \textit{“quality”} in AD, particularly for full-length video, remains an open challenge. To ensure that GenAD produces meaningful and usable descriptions, we plan to establish more comprehensive definitions of quality and advance methodologies for conducting evaluations more efficiently.

Another key direction is emotional speech. Evidence suggests that BLV audiences consistently prefer human narration, which offers clarity and emotional depth that synthetic voices often fail to provide \cite{Fernandez-Torne2015TheCatalan, Walczak2018VocalPresence, RNIBRNIB}. We plan to integrate more expressive TTS voices into both GenAD and AdaptAD could help move beyond functional accessibility toward richer, more engaging experiences.

\section{Conclusion}

This paper introduced ADx3, a collaborative workflow that integrates AI-generated baselines, human-in-the-loop refinement, and adaptive user control to support high quality accessible audio description at scale. Our evaluation with accessibility experts demonstrated that while VLMs, when guided by accessibility-focused prompting—can generate \textit{“good”} descriptions, challenges remain in prioritization, timing, and delivery. RefineAD and AdaptAD address these gaps by enabling editorial nuance and personalizing narration to diverse BLV preferences. 

We emphasize that ADx3 is a formative contribution. The framework illustrates how automation, human editing, and interactive queries can be integrated into a workflow that support meaningful and more adaptive audio description, but full validation with BLV audiences remains essential. Future work will advance this study through expanded datasets and large-scale evaluations, involving accessibility experts, professional describers, novice describers, and, most importantly, BLV users in sustained longitudinal studies. By combining scalable automation with inclusive collaboration, ADx3 positions audio description not as a fixed product but as a dynamic process, expanding access to digital media while centering the agency and experience of BLV audiences.

\begingroup\let\clearpage\relax
\appendix

\section{GenAD Prompt Templates}

This section contains the full prompt templates used in our system, including initial guideline setup, generation, optimization, and refinement prompts.

\subsection{Guidelines Prompt}
\begin{figure}[H]
\centering
\begin{lstlisting}
    guidelines = """
    AUDIO DESCRIPTION GUIDELINES:
    - Describe what you see in a concise, factual manner.
    - Always read on-screen text exactly as it appears.
    - Be factual, objective, and precise in your descriptions.
    - Use proper terminology and names from the context when possible.
    - Match the tone and mood of the video.
    - Do not over-describe - less is more.
    - Do not interpret or editorialize about what you see.
    - Do not give away surprises before they happen.
    
    IMPORTANT CHARACTER IDENTIFICATION:
    - When you recognize a character from the context, ALWAYS use their specific name.
    - Before each scene, carefully review context to identify all named characters.
    - Use the most specific identification possible based on the context information.
    """
    
    guidelines_prompt = """
    You are a professional audio describer following these guidelines:
    
    {guidelines}
    
    Do you understand these guidelines? Respond with "YES" and a brief confirmation.
"""
\end{lstlisting}
\caption{Guidelines prompt for VLM initialization}
\end{figure}

\subsection{Video Generation Prompt}
\begin{figure}[H]
\centering
\begin{lstlisting}
    prompt = """
    SCENE DURATION: {scene_duration:.2f} seconds
    
    CONTEXT:
    {context}
    
    You are analyzing a video scene. Identify specific characters, locations, and any important elements mentioned in the context.
    
    First, generate a JSON array of Text on Screen events.
    Text Events ("type": "Text on Screen"):
    - Capture ALL visible on-screen text.
    - DO NOT include transcript or dialogue.
    - CRITICAL: For each text event, include the EXACT `start_time` in seconds when the text appears.
    
    INCLUDE:
    - Titles, headings, names
    - Informational text
    - Important dates or events
    
    EXCLUDE:
    - Brand logos and watermarks
    - Network logos
    - Social media handles
    - Copyright notices
    
    Second, generate a JSON array of Visual events.
    - Provide contextually rich visual description of the scene using concise wording
    - Describe each action in detail
    - ALWAYS use specific character names from context (not "person" or "woman")
    - Focus on key actions, settings, objects that aren't mentioned in previous description
    - Include clear start times for each visual event
    - DO NOT describe Text on Screen
    
    RULES:
    - Format the output as a JSON array. Each object should include:
      - `start_time` (in seconds)
      - `type` ("Text on Screen" or "Visual")
      - `text` (description or on-screen text)
    """
\end{lstlisting}
\caption{Prompt for generating scene-level descriptions and text events}
\end{figure}

\subsection{Inline Optimization Prompt}
\begin{figure}[H]
\centering
\begin{lstlisting}
    prompt = """
    You are optimizing a set of visual descriptions for a video.
    ORIGINAL DESCRIPTIONS: "{combined_text}"
    
    AVAILABLE TIME: {available_duration:.2f} seconds
    
    TASK:
    Combine and condense these descriptions to fit within {available_duration:.2f} seconds.
    
    GUIDELINES:
    - Create a coherent, flowing description
    - Maintain the action order and use concise but natural language
    - Ensure the final description MUST be spoken within the time limit
    
    OUTPUT FORMAT:
    Provide only the optimized description text, without explanations.
    """
\end{lstlisting}
\caption{Prompt for inline description optimization}
\end{figure}

\subsection{Retry Optimization Prompt}
\begin{figure}[H]
\centering
\begin{lstlisting}
    retry_prompt = """
    You are optimizing multiple visual descriptions for a video.
    PREVIOUS ATTEMPT: "{optimized_text}"
    
    This description takes {tts_duration:.2f} seconds to speak, but you only have {available_duration:.2f} seconds available.
    You need to reduce it by {tts_duration - available_duration:.2f} seconds.
    
    TASK:
    Create a SHORTER version that MUST fit within the time limit.
    
    GUIDELINES:
    - Keep the most critical visual elements and eliminate redundant details
    - Use concise but natural language
    
    OUTPUT FORMAT:
    Provide only the shortened description, nothing else.
    """
\end{lstlisting}
\caption{Prompt for retrying a failed inline optimization}
\end{figure}

\subsection{Extended Description Filtering Prompt}
\begin{figure}[H]
\centering
\begin{lstlisting}
    prompt = """
    You are an accessibility expert selecting ONE visual description per scene to convert to audio description for blind and low-vision users.
    
    CONTEXT:
    - AD should be minimal and only interrupt audio when necessary
    - Spoken transcript is the primary information source
    - Use AD only when critical visual info is missing from the audio
    
    INPUT:
    CURRENT SCENE TRANSCRIPT:
    {transcript_text}
    
    CUMULATIVE TRANSCRIPT:
    {cumulative_transcript}
    
    CUMULATIVE DESCRIPTION:
    {previous_desc_text}
    
    VISUAL DESCRIPTIONS TO EVALUATE:
    {clip['text']}
    
    EVALUATION CRITERIA:
    Include a description (necessary = true) if it meets any of these:
    - Conveys visual details not in the audio
    - Describes important silent actions or key events
    - Identifies characters or locations that are otherwise ambiguous
    - Introduces a novel or distinct visual element
    - Notes scene changes or unannounced time jumps
    
    OUTPUT FORMAT:
    Return a JSON array where item has "necessary": true:
    - "id": index of the description
    - "necessary": true or false
    - "reason": short explanation
    """
\end{lstlisting}
\caption{Prompt for filtering extended AD candidates}
\end{figure}

\subsection{How-To Scene Merging Prompt}
\begin{figure}[H]
\centering
\begin{lstlisting}
    prompt = """
    TASK:
    Combine the text on screen and visual elements into ONE coherent description of the scene, then shorten it to fit within the time limit.
    
    IMPORTANT:
    - Preserve and include ALL measurements exactly
    - The final description MUST fit within {available_scene_time:.2f} seconds
    """
\end{lstlisting}
\caption{Prompt for merging and compressing instructional descriptions}
\end{figure}

\section{AdaptAD Prompt Templates}
\subsection{Key D}
\begin{figure}[H]
\centering
\begin{lstlisting}
    query == "describe the scene":
    prompt = f"""VIDEO SCENE CONTEXT:
        {scene_info_text}
        USER QUERY: {query}
        IMPORTANT: You are looking at two frames from the video.
        - The first frame is a keyframe captured near timestamp {keyframe_time:.2f}s.
        - The second frame is the exact frame captured at timestamp {exact_time:.2f}s.  
        """
\end{lstlisting}
\caption{Prompt when user press key \textit{D} to request additional descriptions of the scene}
\end{figure}

\subsection{Key Q}
\begin{figure}[H]
\centering
\begin{lstlisting}
    prompt = f"""VIDEO SCENE CONTEXT:
        {scene_info_text}
        USER QUERY: {query}
        IMPORTANT: You are looking at two frames from the video.
        - The first frame is a keyframe captured near timestamp {keyframe_time:.2f}s.
        - The second frame is the exact frame captured at timestamp {exact_time:.2f}s. 
        Use context to provide a contextually rich answer to the {query} in one very concise sentence.
        When the query is a "Why" question, your answer must include the specific cause-and-effect details 
        (identifying who was involved, when the event occurred, and what happened) that directly answer the query.
        If the question is Where, be specific with the setting.
        Do NOT add external details outside of scope.
        Do NOT refer to frame numbers or timestamps.
        """
\end{lstlisting}
\caption{Prompt when user press key \textit{D} to request additional descriptions of the scene}
\end{figure}

\section{Multi-Dimensional Assessment Model for Audio Description}

This appendix section reproduces the complete documentation provided to expert raters. 

\textbf{Overview}: 
This model evaluates both human- and AI-generated audio description for YouTube videos across two main dimensions: \textbf{Content} (five criteria) and \textbf{Formatting} (two criteria).   

\subsection*{I. Content}

\textit{(1) Accurate — Error-Free Content}

\textbf{Definition:} Description provides error-free visual information with correct identification of what's actually happening. No factual mistakes or misleading information.  

\textbf{Evaluation Criteria (1–5):}  
\begin{itemize}
    \item[5:] All visual elements are factually correct. No errors in describing what's actually happening. Perfect factual accuracy.
    \item[4:] Mostly factually correct with minor errors that don't mislead. Generally accurate descriptions.  
    \item[3:] Generally factually correct but with some noticeable errors. Mostly accurate with some mistakes. 
    \item[2:] Multiple factual errors that mislead about what's happening. Poor accuracy in descriptions.
    \item[1:] Major factual errors or completely incorrect information. Fails to accurately describe what's happening.
\end{itemize}

\textit{(2) Prioritized — Context \& Inference}

\textbf{Definition:} The description achieves optimal prioritization by selecting details based on their contextual significance and inferential value. For example, the description prioritizes contextually-rich details over generic descriptions such as "Harry runs into the forest" vs. "a boy runs into the forest", and makes reasonable inferences, such as "a boy in a soccer uniform" vs "a boy in red jersey and tall socks". 

\textbf{Evaluation Criteria (1–5):}  
\begin{itemize}
    \item[5:] Just right balance - perfect prioritization on most significant elements for understanding. Chooses contextually relevant details and appropriate spatial information.
    \item[4:] Good prioritization but not perfect - either slightly too generic or slightly excessive. Generally good choices about what to include  
    \item[3:] Adequate  prioritization but noticeable imbalance - either missing some important details or including some unnecessary information. 
    \item[2:] Poor prioritization - either incomplete important information or includes too many unimportant details. Poor choices about what matters.
    \item[1:] Major problems - either major gaps in important information or describes everything including unimportant elements. No clear prioritization on what's significant. 
\end{itemize}

\textit{(3) Appropriate — Audience \& Purpose Alignment}

\textbf{Definition:} The language, level of detail, and style of the description should suit the type of content and the intended audience experiences. For example, for entertainment videos, the description should enhance enjoyment, for educational videos, it should support understanding, and instructional videos should enable viewers to follow or replicate the steps shown.

\textbf{Evaluation Criteria (1–5):}  
\begin{itemize}
    \item[5:] Perfect alignment - language and detail level expertly matched to both audience capabilities and content purpose. Description fully supports intended experience.  
    \item[4:] Good alignment with minor mismatches - generally appropriate for audience and purpose but occasional lapses in tone, complexity, or focus.  
    \item[3:] Adequate alignment but noticeable disconnects - partially serves audience and purpose but inconsistent in matching language level or functional needs. 
    \item[2:] Poor alignment - frequently uses inappropriate language for the audience or fails to support content purpose. Description often works against intended goals.  
    \item[1:] Complete misalignment - language and approach entirely unsuited to the audience and/or actively undermines content purpose. No apparent consideration of who will use this or why.  
\end{itemize}

\textit{(4) Consistent — Consistency \& Coherence}

\textbf{Definition:} The description maintains consistent terminology, style, and tone, supporting a coherent and unified narrative throughout the video.  

\textbf{Evaluation Criteria (1–5):}  
\begin{itemize}
    \item[5:] Fully consistent in terminology and style. Narrative flows smoothly and coherently.
    \item[4:] Mostly consistent with minor variations. The narrative remains generally coherent.  
    \item[3:] Adequate consistency, but some noticeable shifts in terminology or style.  
    \item[2:] Frequent inconsistencies in word choice or tone. The narrative becomes difficult to follow.  
    \item[1:] No consistency maintained. The narrative is disjointed or incoherent.  
\end{itemize}

\textit{(5) Equal — Objectivity \& Non-Interpretation}

\textbf{Definition:} The description ensures equal access by being objective and without personal interpretation, bias, or unnecessary commentary.  

\textbf{Evaluation Criteria (1–5):}  
\begin{itemize}
    \item[5:] Completely objective. No personal interpretation. Appropriate descriptive language without editorial comment.  
    \item[4:] Generally objective with rare minor interpretive moments.  
    \item[3:] Mostly objective but some unnecessary interpretation present.  
    \item[2:] Frequent interpretive language. Some bias evident in descriptions.  
    \item[1:] Highly interpretive and biased. Significant personal commentary interferes with equal access.  
\end{itemize}

\subsection*{II. Formatting}

\textit{(1) Strategic Use of Description Method (Inline vs. Extended)}

\textbf{Definition:} The description makes effective choices between inline and extended description methods based on content characteristics.

\textbf{Inline }description is the standard and preferred method when:
\begin{itemize}
    \item Sufficient natural pauses exist within original content timing \cite{NationalCenterforAccessibleMedia2017AccessibleGuidelines} 
    \item Visual content can be adequately described within available audio gaps\cite{3PlayMedia2020AudioGuidelines}
\end{itemize}

    \textbf{Extended} description is appropriate when:
\begin{itemize}
    \item Text-heavy videos, like recordings of slideshows or lectures \cite{3PlayMedia2020AudioGuidelines}
    \item Dialogue-heavy videos, as audio description shouldn't drown out what people are saying \cite{3PlayMedia2020AudioGuidelines}
    \item Noisy videos containing important music or sound, as audio description could detract from these elements \cite{3PlayMedia2020AudioGuidelines}
    \item Videos with short cuts and/or extremely detailed frames where standard description would be incomplete by the next cut \cite{3PlayMedia2020AudioGuidelines}
    \item Essential visual information cannot adequately fit within available natural pauses: "If no such pause exists, you must insert an extended description at that point \cite{NationalCenterforAccessibleMedia2017AccessibleGuidelines}
\end{itemize}

\textbf{Evaluation Criteria (1–5):}  
\begin{itemize}
    \item[5:] Perfect method selection - consistently chooses inline for content with adequate pauses, extended only when absolutely necessary based on professional criteria.
    \item[4:] Good method selection with occasional minor errors - generally appropriate choices with rare unnecessary use of extended description.
    \item[3:] Adequate method selection but some poor choices - sometimes uses extended unnecessarily or misses opportunities when extended is needed.  
    \item[2:] Poor method selection - frequently uses wrong method, either overusing extended description or failing to use it when required.  
    \item[1:] Severe method selection issues - no understanding of when to use inline vs. extended based on professional standards.  
\end{itemize}

\textit{(2) Timing \& Placement}

\textbf{Definition:} Appropriate timing of description placement relative to visual content and audio elements based on established accessibility standards. Timing standards for both description methods as follows: 
\begin{itemize}
    \item No interruption of important dialogue or essential sound effect \cite{2024DescribedDCMP}.  
    \item Insert descriptions at natural points in the timeline - don't cut off speakers mid-word, but take advantage of brief pauses. Even pauses between words or sentences suffice as long as the description is not out of context \cite{NationalCenterforAccessibleMedia2017AccessibleGuidelines}.  
    \item Place descriptions as close to the visual action as possible \cite{3PlayMedia2020AudioGuidelines}.
    \item Pre-description is allowed: descriptions may be inserted "slightly before the action occurs on screen" if it clarifies the situation \cite{NationalCenterforAccessibleMedia2017AccessibleGuidelines}.
\end{itemize}

\textbf{Evaluation Criteria (1–5):}  
\begin{itemize}
    \item[5:] Optimal timing - descriptions placed during natural pauses close to the visual action without interrupting essential audio.  
    \item[4:] Occasionally poor timing - generally good placement but sometimes descriptions are too early, too late, or slightly overlap important audio.  
    \item[3:] Noticeable timing issues - descriptions poorly timed relative to visual content, some interference with dialogue.  
    \item[2:] Poor timing - descriptions often mistimed, frequently interrupting dialogue or placed too far from relevant action.  
    \item[1:] Severe timing issues - consistently poor timing that disrupts content flow and interferes with essential audio.  
\end{itemize}

\endgroup

\bibliographystyle{ACM-Reference-Format}
\bibliography{references-new}

\end{document}